\documentclass[preprint,12pt]{elsarticle}
\pdfoutput=1 

\usepackage{hyperref}
\usepackage{amssymb}
\usepackage{amsthm}
\usepackage{amstext}
\usepackage{amsmath}
\usepackage{breqn}
\usepackage{verbatim}
\usepackage{epstopdf}
\setkeys{breqn}{breakdepth={0}}
\usepackage{amsfonts}
\usepackage{bbm}
\usepackage[german,english]{babel}
\usepackage{multirow}
\usepackage{booktabs}

\def\@preprint{\@empty}
\newcommand\preprint[1]{\gdef\@preprint{\hfill #1}}

\newcommand{\ra}[1]{\renewcommand{\arraystretch}{#1}}

\newcounter{bla}

\journal{Computer Physics Communications}

\usepackage{enumitem}
\usepackage{algorithm}
\usepackage{algorithmic}

\usepackage{tikz}
\usetikzlibrary{positioning}
\usetikzlibrary{arrows.meta}
\usetikzlibrary{shapes}
\usepackage{hyperref}
\usepackage{listings}
\usepackage{xcolor}
\usepackage{todonotes}
\definecolor{codegreen}{rgb}{0,0.6,0}
\definecolor{codegray}{rgb}{0.5,0.5,0.5}
\definecolor{codepurple}{rgb}{0.58,0,0.82}
\definecolor{backcolour}{rgb}{0.93,0.93,0.93}

\theoremstyle{plain}
\newtheorem{thm}{Theorem}[section]

\theoremstyle{definition}
\newtheorem{defn}[thm]{Definition}

\newtheorem{rem}[thm]{Remark}

\begin{document} 
	
\preprint{PCFT-21-08, USTC-ICTS-21-08}

\begin{frontmatter}
\title{{\sc pfd-parallel}, a Singular/GPI-Space package for massively parallel multivariate partial fractioning}
\author[add1,add6]{Dominik Bendle}
\ead{dominik.bendle@itwm.fraunhofer.de}

\author[add1]{Janko Boehm}
\ead{boehm@mathematik.uni-kl.de}

\author[add1,add6]{Murray Heymann}
\ead{heymann@mathematik.uni-kl.de}

\author[add3,add2]{Rourou Ma}
\ead{marr16@lzu.edu.cn}

\author[add6]{Mirko Rahn}
\author[add1]{Lukas Ristau}
\author[add1]{Marcel Wittmann}
\ead{mwittman@rhrk.uni-kl.de}

\author[add3,add4]{Zihao Wu}
\ead{wuzihao@mail.ustc.edu.cn}
\author[add3,add4,add5]{Yang Zhang}
\ead{yzhphy@ustc.edu.cn}

\address[add1]{Department of Mathematics, Technische Universit\"at Kaiserslautern, 67663 Kaiserslautern, Germany}
\address[add2]{Cuiying Honors College, Lanzhou University, Lanzhou, Gansu 730000, China}
\address[add3]{Interdisciplinary Center for Theoretical Study, University of Science and Technology of China, Hefei, Anhui 230026, China}
\address[add4]{Peng Huanwu Center for Fundamental Theory, Hefei, Anhui 230026, China}
\address[add5]{Max-Planck-Institut f\"ur Physik,
  Werner-Heisenberg-Institut, D-80805, M\"unchen, Germany}
\address[add6]{Fraunhofer Institute for Industrial Mathematics
 (ITWM), Fraunhofer-Platz 1, 67663 Kaiserslautern, Germany}

\begin{abstract}
Multivariate partial fractioning is a powerful tool for simplifying rational
function coefficients in scattering amplitude
computations. Since current research problems lead to large sets of complicated rational
functions, performance of the partial fractioning as well as size of the obtained expressions are a prime concern. We develop a large scale parallel framework for multivariate partial fractioning, which implements and combines an improved version of Leinartas' algorithm and the {\sc
    MultivariateApart} algorithm.  Our approach relies only on open source software. It combines parallelism over the different rational function
  coefficients with parallelism for individual expressions. The implementation is based on the 
  \textsc{Singular}/\textsc{GPI-Space framework} for massively parallel computer algebra, which formulates parallel algorithms in terms of Petri nets. The modular nature of this approach allows for easy incorporation of future algorithmic developments into our package. 
 We demonstrate the performance of our framework by simplifying expressions arising from current multiloop scattering amplitude problems.
\end{abstract}


\end{frontmatter}

{\bf PROGRAM SUMMARY}

\begin{small}
\noindent
{\em Manuscript Title:} {\sc pfd-parallel}, a Singular/GPI-Space package for massively parallel multivariate partial fractioning\\
{\em Authors:} Dominik Bendle, Janko Boehm, Murray Heymann, Rourou Ma,
Mirko Rahn, Lukas Ristau, Marcel Wittmann, Zihao Wu, Yang Zhang\\
{\em Program Title:}   pfd-parallel                                       \\
{\em CPC Library link to program files:} (to be added by Technical Editor) \\
{\em Developer's repository link:} https://github.com/singular-gpispace/pfd-parallel\\
{\em Code Ocean capsule:} (to be added by Technical Editor)\\
{\em Licensing provisions(please choose one):} GPLv3  \\
{\em Programming language:}         Singular language, GPI-Space Petri net                         \\
{\em Computer(s) for which the program has been designed:} from personal computer to HPC cluster\\
{\em Operating system(s) for which the program has been designed:} Linux\\
{\em Supplementary material:} none                                \\
{\em Nature of problem:} 
In scattering amplitude computation, we often encounter complicated
rational functions, which may be simplified using the multivariate partial fraction decomposition. With the consideration of the increasing complexity of the problems, an efficient implementation of such a method is needed. 
\\
{\em Solution method:} We present the package {\sc pfd-parallel},
which is a large-scale parallelized framework  for multivariate
partial fractioning. Our package relies only on open source software. It combines different algorithms and provides parallelization based on the {\sc Singular/GPI-Space
framework} \cite{5,2,3}. The package combines both the improved Leinartars'
algorithm \cite{1}, as well as the {\sc MultivariateApart} algorithm \cite{4}, and combines parallelism over the different rational function
  coefficients and parallelism for individual expressions. Using this approach cutting-edge computations can be handled in an efficient way.

{\em Additional comments including restrictions and unusual features (approx. 50-250 words):}
The software, including all dependencies like Singular, GPI-2, GPI-Space, and the Singular/GPI-Space framework, is distributed via the supercomputing package manager Spack, which allows for convenient installation of scientific software, in particular in HPC environments. The code has been tested on Centos 7 and 8, Ubuntu 18.04 LTS and 20.04 LTS.\\
   \\

\end{small}

\section{Introduction}
\label{sec:intro}
Higher order perturbative quantum field theory correction is a very
important theoretical tool in high energy physics. The higher order
correction is usually obtained by the computation of
multi-loop scattering amplitudes. One difficulty in scattering amplitude
computations arises from the complicated rational function coefficients in
terms of the kinematic variables. Complicated
rational functions are difficult to handle algebraically to evaluate numerically. 

Algorithmic multivariate partial fractioning, originating from the mathematical work
of Leinartas \cite{leinartas1978factorization, raichev2012leinartas}, is a powerful tool which can simplify many kinds of
rational function expressions in scattering amplitude computations. See the recent improved Leinartas' algorhtm in ref.~\cite{Boehm:2020ijp}. On the other hand, the  {\sc MultivariateApart}~\cite{Heller:2021qkz} algorithm, based on computation algebraic geometry, is significantly different from Leinartas' and has strong performance. Multivariate partial fractioning was used in the algorithm in
\cite{Meyer:2017joq} for finding Feynman 
integrals with uniform transcendental weights~\cite{Henn:2013pwa}, in
simplifying integration-by-parts (IBP) reduction
coefficients~\cite{Boehm:2020ijp}, and, with a multitude of applications, in simplifying the
rational function coefficients in the final expressions of multiloop
scattering amplitudes~\cite{Abreu:2019odu,Agarwal:2021grm,
  Agarwal:2021vdh, Badger:2021nhg,Badger:2021imn, Badger:2021ega,Abreu:2021asb,Badger:2022ncb}. 

In multiloop amplitude computations, both the number and complexity of rational functions to be handled can be very significant. Therefore the performance
of the multivariate partial fraction implementation is  important.
 A large scale parallelization of multivariate
partial fractioning, over different rational function coefficients is obviously a necessity. Furthermore,  the rational function
coefficients from a scattering amplitude computation typically have quite
uneven size and thus dramatically different running times for the
partial fraction decomposition. A naive parallelization over batches of different rational
functions is then not efficient. For very complicated rational
functions, a parallelization of the decomposition of individual functions would be helpful.
Realizing both types of parallelism simultaneously in an efficient way requires to model a non-trivial workflow.

 There are several publicly available programs  for applying multivariate partial fraction methods, like {\sc pfd}~\cite{Boehm:2020ijp}, {\sc MultivariateApart}~\cite{Heller:2021qkz},
as well as private codes. In this paper, we present a large-scale
parallel implementation of
the two algorithms in \cite{Boehm:2020ijp}
and \cite{Heller:2021qkz}, based on the \textsc{Singular}/\textsc{GPI-Space} framework\footnote{\href{https://www.mathematik.uni-kl.de/~boehm/singulargpispace/}{https://www.mathematik.uni-kl.de/\textasciitilde  boehm/singulargpispace/}} introduced in \cite{singgpi}.  Our package is solely based on open source software. It can be conveniently installed  via the widely-used supercomputing package manager \textsc{Spack} \cite{spack}, and is usable on homogeneous and heterogenous computing environments from a personal computer to a high-performance cluster. 

In our approach,  the open source computer algebra system \textsc{Singular} \cite{DGPS} provides efficient tools for handling algebraic structures, and in particular Gr\"obner bases. We rely on the powerful task-based workflow management system \textsc{GPI-Space} \cite{GSPC} to model our parallel algorithm. This system has also been made open source. By formulating the algorithm in  \textsc{GPI-Space} in the language of Petri nets leads to automated parallelization.
 The parallelization is
realized not just on the level of different coefficients, but also {\it within
  the partial fraction computation of individual coefficients}. This feature is
very useful for simplifying rational function coefficients if the sizes of these coefficients are quite uneven, which is typically the case. Due to the
internal parallelization, complicated
coefficients do not dominate any more the run-time. In particular, a very good parallel efficiency is achieved. 

Our package also provides  
convenient strategies to choose different partial fractioning algorithms and types of parallelism based on the sizes of the input functions. The package and the realization of the parallelism are designed with a modular structure in mind, allowing for easy integration of improvements and algorithms which might become available in the future.

To demonstrate the power of our package, we present two examples of
simplifying (1) IBP reduction
coefficients for two-loop five-point
massless Feynman integrals with degree $5$ numerators, and (2) rational functions from two-loop leading colour helicity amplitudes
for $W \gamma + j$ production \cite{Badger:2022ncb}. Both cases can be handled with our package in a highly efficient way, and the
resulting representation of the data is significantly shorter.

This paper is organized in the following way: In Section \ref{sec:algorithm}, we introduce our improved Leinartas' algorithm
equipped with polynomial division and syzygy reduction
computations, for the multivariate partial fraction. The parallel
structures are also mentioned. In Section \ref{PFD}, we discuss our new large-scale
parallel implementation of our partial fraction algorithm based on the
\textsc{Singular}/\textsc{GPI-Space} framework. In Section \ref{sec:pfd_usage}, we provide a note on the installation and a short manual for our program.\footnote{More details can be found in the online manual in the GitHub repository \url{https://github.com/singular-gpispace/pfd-parallel}.} In Section \ref{sec:examples}, we apply our multivariate
partial fractioning framework in the case of our examples (1) and (2), and analyze timings and compression ratios.
We observe that  the IBP coefficients for the two-loop five-point non-planar Feynman integral with
degree $5$ numerators are compressed by more than two orders of magnitudes and are
put into a usable form, moreover, that the amplitude example achieves similar compression ratios. Finally, in Section~\ref{sec:outlook}, we summarize our paper and
provide some outlook.

\section{Partial fraction decomposition}
\label{sec:algorithm}

The algorithm we use to reduce the size of rational functions is an improved version of Leinartas' algorithm for multivariate partial fraction decomposition~\cite{leinartas1978factorization, raichev2012leinartas}. Since our approach for parallelization of the decomposition of individual functions relies on it, we start out with a short account of the improved Leinartas algorithm. For more details refer to Section 3 of our paper~\cite{Boehm:2020ijp}. For the the {\sc MultivariateApart} algorithm, of which we also provide an implementation, we refer to~\cite{Heller:2021qkz}.

Let in the following $K[x_1,\dots,x_d]$ or short $K[\mathbf x]$ be the ring of polynomials over some field $K$ in $d$ variables $\mathbf x=(x_1,\dots,x_d)$ and let $\overline K$ be the algebraic closure of $K$ (e.g. $\overline{\mathbb R} = \mathbb C$).
The goal is to write a rational function $f/g$ ($f,g\in K[\mathbf x]$) where the polynomial 
$g=q_1^{e_1}\cdot\dots\cdot q_m^{e_m}$
factors into many small\footnote{In rational functions arising from IBP reductions most of the denominator factors are of degree 1.} irreducible factors $q_i$, as a sum of functions with ``smaller'' numerators and denominators. The algorithm consists of 3 main steps:

\subsection{Nullstellensatz decomposition}\label{sec:NSSDec}
In the first step of the algorithm we search for relations of the form 
\begin{equation}\label{eq:NSScertificate}
1=h_1q_1^{e_1}+\dots+h_mq_m^{e_m}
\end{equation}
where $h_i\in K[\mathbf x]$ are polynomials. By multiplying (\ref{eq:NSScertificate}) with $f/g$, we get the decomposition
\begin{equation}\label{eq:NSSdecomp}
\frac f g=\frac{f\cdot\sum_{k=1}^mh_kq_k^{e_k}}{\prod_{i=1}^mq_i^{e_i}}=\sum_{k=1}^m\frac{f\cdot h_k}{\prod_{i=1,i\not=k}^mq_i^{e_i}}.
\end{equation}
in which each denominator contains only $m-1$ different irreducible factors. Now we repeat this step with each summand in the decomposition (\ref{eq:NSSdecomp}) until we obtain a sum of rational functions where each denominator contains only factors, that do not admit a relation as in (\ref{eq:NSScertificate}). By Hilbert's weak Nullstellensatz~\cite[Lemma~3.6]{Boehm:2020ijp}, such a relation exists if and only if the polynomials $q_i$ do not have a common zero in $\overline K^d$ and can be computed by calculating a Gr\"obner basis~\cite[Definition~3.3]{Boehm:2020ijp} of the ideal generated by the polynomials $q_i^{e_i}$ \cite[Algorithm~1]{Boehm:2020ijp}. 

\subsection{Algebraic dependence decomposition}\label{sec:algDepDec}
If the polynomials $q_1^{e_1},\dots,q_m^{e_m}$ are algebraically dependent, i.e. there exists a polynomial $p\in K[y_1,\dots,y_m]$ in $m$ variables called an annihilating polynomial for $q_1^{e_1},\dots,q_m^{e_m}$, such that $p(q_1^{e_1},\dots,q_m^{e_m})=0 \in K[\mathbf x]$, then we can use this equation to derive a decomposition similar to (\ref{eq:NSSdecomp}).
For this, write 
\begin{align}\label{eq:alpha}
p=c_\alpha\mathbf y^\alpha+\sum_{\substack{\beta\in\mathbb N^m\\ \deg(p)\ge|\beta|\ge|\alpha|}} c_\beta\mathbf y^\beta\qquad (c_\alpha,c_\beta\in K, c_\alpha\not=0)
\end{align}
such that $c_\alpha\mathbf y^\alpha$ is one of the terms of smallest degree (using multi-indices $\beta\in\mathbb N^m$, so $\mathbf y^\beta=y_1^{\beta_1}\cdot\dots\cdot y_m^{\beta_m}$ and $\deg(\mathbf y^\beta)=|\beta|=\beta_1+\dots+\beta_m$). Writing $\mathbf q$ for the vector $(q_1^{e_1},\dots,q_m^{e_m})$, it holds\begin{align}\label{eq:algDependDecomp}
0 = p(\mathbf q)\quad &\Leftrightarrow\quad c_\alpha\mathbf q^\alpha =-\sum_{\beta} c_\beta\mathbf q^\beta\nonumber\\
&\Leftrightarrow\quad 1=-\sum_{\beta} \frac{c_\beta\mathbf q^\beta}{c_\alpha\mathbf q^\alpha}=-\sum_{\beta} \frac{c_\beta}{c_\alpha}\prod_{i=1}^m\frac{q_i^{e_i\beta_i}}{q_i^{e_i\alpha_i}}\nonumber\\
&\Rightarrow\quad \frac f g=-\sum_{\beta} \frac{c_\beta}{c_\alpha}f\prod_{i=1}^m\frac{q_i^{e_i\beta_i}}{q_i^{e_i(\alpha_i+1)}}
\end{align}
and since $\mathbf y^\alpha$ has minimal degree, for each $\beta$ in the sum in Equation (\ref{eq:algDependDecomp}) it holds  $\beta_i\ge\alpha_i+1$ \linebreak for at least one index $i$, i.e. the factor $q_i$ does not appear in the denominator of the corresponding term and thus the denominators of the rational functions in the decomposition each have at most $m-1$ different irreducible factors.

As with the Nullstellensatz decomposition, this step is repeated with each summand in (\ref{eq:algDependDecomp}). This leads to a decomposition where each denominator contains only algebraically independent factors $q_i$, since it can be shown~\cite[Corollary~3.8]{Boehm:2020ijp}, that polynomials $q_1,\dots,q_m$ are algebraically dependent if and only if $q_1^{e_1},\dots,q_m^{e_m}$ are (for any $e_i\in\mathbb N_{\ge 1}$).

The problem of calculating annihilating polynomials can be reduced to the computation of the Gr\"obner basis of a certain ideal~\cite[Lemma~3.9,~3.10 and Algorithm~2]{Boehm:2020ijp}. But there is a simpler way of determining beforehand, whether an annihilating polynomial exists: The Jacobian criterion states, that a set of polynomials $\left\{g_1,\dots,g_m\right\}$ is algebraically independent if an only if the Jacobian $m\times d$-matrix of polynomials $\left(\frac{\partial g_i}{\partial x_j}\right)_{i,j}$ has full row rank over the field $K(\mathbf x)$ of rational functions~\cite[Lemma~3.7]{Boehm:2020ijp}. From this it also follows, that after the algebraic dependence decomposition, in each denominator the number of \textit{different} irreducible factors is at most $d$ (the number of variables) since an $m\times d$-matrix with $m>d$ cannot have full row rank and thus any $d+1$ polynomials are algebraically dependent.

\subsection{Numerator decomposition}\label{sec:numDec}
Note that in the previous two steps, the denominators become simpler (with respect to the number of different factors in their factorisation), but the numerators do not. In (\ref{eq:NSSdecomp}) and (\ref{eq:algDependDecomp}) the original numerator $f$ still appears in each summand. To also shorten the numerators, it makes sense to do a division with remainder by the factors in the denominator. For a rational function $f/g$ with factorisation $g=q_1^{e_1}\cdot\dots\cdot q_m^{e_m}$ as above we can calculate a division expression
\begin{align}\label{eq:divrem}
f&=r+\sum_{k=1}^ma_kq_k
\end{align}
where $a_i\in K[\mathbf x]$ are polynomials and $r\in K[\mathbf x]$ is a ``small'' remainder. More precisely, by making use of a Gr\"obner basis of the ideal $I=\left<q_1,\dots,q_m\right>$ generated by the irreducible factors $q_i$ we can ensure, that each term of the polynomial $r$ is not divisible by the lead term of \textit{any} element of $I$ \cite[Definition~3.4 and Algorithm~3]{Boehm:2020ijp}. We say, that $r$ is ``reduced'' with respect to~$I$. 
Simply multiplying (\ref{eq:divrem}) by $f/g$ yields
\begin{align}\label{eq:numeratorDecomp}
\frac f g &=\frac{r}{\prod_{i=1}^m q_i^{e_i}} + \sum_{k=1}^m \frac{a_k}{q_k^{(e_k-1)}\prod_{i=1,i\not=k}^mq_i^{e_i}}
\end{align}
where the first term has a particularly small numerator and in each of the other terms one of the factors in the denominator cancels. Thus repeatedly applying this decomposition step results in a sum of rational functions where the numerator of any function is reduced (as defined above) with respect to (the ideal generated by the irreducible factors of) its denominator.

Note that the division expression (\ref{eq:divrem}) depends on the choice of a monomial ordering, i.e. a total ordering on the set $\left\{\mathbf x^\alpha\middle|\alpha\in\mathbb N^d\right\}$ of monomials, that is compatible with multiplication \cite[Definition~3.2]{Boehm:2020ijp}. This ordering is needed to define the ``lead term'' of a multivariate polynomial in the division-with-remainder algorithm. In our \textsc{Singular} implementation we used the graded reverse lexicographic ordering \cite[(3.5)]{Boehm:2020ijp} which sorts first by the degree of the monomial.

\subsection{The resulting algorithm}\label{sec:algo}
If we do the Nullstellensatz decomposition, the algebraic dependence decomposition and the numerator decomposition one after the other, we obtain a sum of rational functions where each summand is of the form
\begin{equation}\label{eq:summand}
\frac{f_S}{\prod_{i\in S}q_i^{b_i}}
\end{equation}
where $S\subseteq\{1,\dots,m\}$ is some set of indices, $b_i\in\mathbb N$, $f_S\in K[\mathbf x]$ and by the above
\begin{enumerate}[label=(\arabic*)]
	\item \label{cond1} the polynomials $\left\{q_i\middle|i\in S\right\}$ have a common zero in $\overline K^d$,
	\item \label{cond2} the polynomials $\left\{q_i\middle|i\in S\right\}$ are algebraically independent,
	\item \label{cond3} $f_S$ is reduced with respect to the ideal $\left<q_i\middle|i\in S\right>\subseteq K[\mathbf x]$. 
\end{enumerate}
(This is Theorem~3.5 in \cite{Boehm:2020ijp}.)

Since in practice, the computation of annihilating polynomials can become very slow if the degrees of the polynomials $q_i^{e_i}$ get too big, we make the following two modifications to the algorithm.
Firstly, before the algebraic dependence decomposition we insert a short version of the numerator decomposition step described in \ref{sec:numDec}, which only decomposes further if the remainder $r$ in (\ref{eq:divrem}) is zero. This eliminates some of the denominator factors before going into the more complicated algebraic dependence decomposition (see also Remark~1.2 and Algorithm~4 in \cite{Boehm:2020ijp}). Secondly, the algebraic dependence decomposition step itself can be changed to using an annihilating polynomial for $q_1,\dots,q_m$ rather than $q_1^{e_1},\dots,q_m^{e_m}$. Instead of (\ref{eq:algDependDecomp}), we then get
\begin{equation}\label{eq:simple_algDependDecomp}
\frac f g=-\sum_{\beta} \frac{c_\beta}{c_\alpha}f\prod_{i=1}^m\frac{q_i^{\beta_i}}{q_i^{\alpha_i+e_i}}.
\end{equation}
Now the number of different irreducible denominator factors does not have to decrease in every step, since it is possible, that $\beta_i<\alpha_i+e_i$ for all $i$. However, if in (\ref{eq:alpha}) we always choose $\alpha$ minimal with respect to the graded reverse lexicographic ordering, it can be shown, that the algorithm still terminates (see Remark~1.3 in \cite{Boehm:2020ijp}).

In our implementation of the final algorithm \cite[Algorithm~5]{Boehm:2020ijp}, we make use of the computer algebra system \textsc{Singular}, which provides efficient algorithms for Gr\"obner basis computations as well as polynomial factorization and division with remainder.

\subsection{A simple example}
To demonstrate the algorithm described in \ref{sec:algo} consider the rational function
\begin{equation}
\frac f g=\frac{x_1+x_2}{x_1x_2(x_2+1)x_3(x_1-x_3)}\in\mathbb R[x_1,x_2,x_3]
\end{equation}
with $m=5$ denominator factors $q_1=x_1,\; q_2=x_2,\; q_3=x_2+1,\; q_4=x_3,\; q_5=x_1-x_3$.

In the first step (\emph{Nullstellensatz decomposition}) we observe, that $q_2$ and $q_3$ have no common zeros and find the relation $1=1\cdot q_3+(-1)\cdot q_2$. Multiplying with $f/g$ yields
\begin{equation}\label{eq:nssdec}
\frac f g = \frac{x_1+x_2}{q_1q_2q_4q_5} + \frac{-x_1-x_2}{q_1q_3q_4q_5}.
\end{equation}
Now $q_1,q_2,q_4,q_5$ have the common zero $x_1=x_2=x_3=0$ and also $q_1,q_3,q_4,q_5$ have a common zero, namely $x_1=x_3=0,\; x_2=-1$. So condition \ref{cond1} is fulfilled.

In the \emph{short numerator decomposition} described in Section \ref{sec:algo} we see that only the first numerator $x_1+x_2$ has remainder $0$ when dividing by the denominator factors $q_1,q_2,q_4,q_5$: $x_1+x_2=1\cdot q_1+1\cdot q_2+0$. Thus in the first term, we can cancel $q_1$ and $q_2$ respectively:
\begin{equation}\label{eq:sndec}
\frac f g = \frac{1}{q_2q_4q_5} + \frac{1}{q_1q_4q_5} + \frac{-x_1-x_2}{q_1q_3q_4q_5}.
\end{equation}

Next, in the \emph{algebraic dependence decomposition}, the factors in the second and third denominator of (\ref{eq:sndec}) are found to be algebraically dependent, since $0=q_1-q_4-q_5$. Thus $1 = \frac{q_1}{q_5}+\frac{-q_4}{q_5}$ and multiplying this to the second and third term gives the decomposition
\begin{equation}\label{eq:algdepdec}
\frac f g = \frac{1}{q_2q_4q_5} + \frac{1}{q_4q_5^2} + \frac{-1}{q_1q_5^2} + \frac{-x_1-x_2}{q_3q_4q_5^2} + \frac{x_1+x_2}{q_1q_3q_5^2}
\end{equation} 
in which all denominators consist of algebraically independent factors, so \ref{cond2} is fulfilled.

Finally, in the \emph{numerator decomposition}, the first three numerators are already reduced with respect to the denominator factors since they are just $\pm1$. For the fourth numerator we get the division expression $-x_1-x_2=(-1)\cdot q_3+(-1)\cdot q_4+(-1)\cdot q_5+1$ and for the fifth numerator it holds $x_1+x_2=1\cdot q_1+1\cdot q_3+(-1)$. Substituting this into (\ref{eq:algdepdec}) yields
\begin{align}\label{eq:ndec}
\frac f g &= \frac{1}{q_2q_4q_5} + \frac{1}{q_4q_5^2} + \frac{-1}{q_1q_5^2} 
+ \frac{-1}{q_4q_5^2} + \frac{-1}{q_3q_5^2}+ \frac{-1}{q_3q_4q_5} + \frac{1}{q_3q_4q_5^2}
+ \frac{1}{q_3q_5^2} \nonumber\\ &+ \frac{1}{q_1q_5^2} + \frac{-1}{q_1q_3q_5^2}\nonumber\\
&= \frac{1}{q_2q_4q_5} + \frac{-1}{q_3q_4q_5} + \frac{1}{q_3q_4q_5^2} + \frac{-1}{q_1q_3q_5^2}
\end{align} 
which now also satisfies condition \ref{cond3}. Note, that for this simple example the partial fraction decomposition (\ref{eq:ndec}) does not seem ``shorter'' than the original fraction $f/g$. However, as described in the following, for the large functions occurring in amplitude and IBP problems, this algorithm can reduce the size of rational functions by factors of more than $100$.

\subsection{Parallel structures in the algorithm}\label{sec:parallel}
To reduce the runtime, it is of course possible to run the partial fractioning algorithm in parallel for all functions occurring in the problem under consideration. However, it is often the case that a small number of large functions dominate and hence determine the total runtime.
Therefore it makes sense to also parallelize the PFD algorithm itself, at least for the most complicated functions in the given problem set. 

In the implementation of the algorithm described above we work with a list $D$ of terms representing a sum and start with $D$ containing only the input rational function as single entry. Then each step of the algorithm consists of decomposing all terms in $D$ individually into a sum/list of terms itself 
(as in equations (\ref{eq:NSSdecomp}), (\ref{eq:numeratorDecomp}) and (\ref{eq:simple_algDependDecomp}))
and then replacing $D$ by the concatenation of these lists as well as merging terms that have the same denominator (see lines 4, 7, 10, 13 in \cite[Algorithm~5]{Boehm:2020ijp}). The processing of summands can be done in parallel over all elements of $D$, leading to \emph{Fork/Join}  patterns, which can be used for parallelization. We discuss parallelism in detail in the subsequent sections.


\section{Parallel computing using the Singular/GPI-Space framework}
\label{PFD}

The implementation of our RREF algorithm and the partial fraction decomposition algorithm is part of the
\textsc{Singular}/\textsc{GPI-Space} framework project for
massively parallel computations
in computer algebra \cite{singgpi}. This framework combines the
open source computer algebra system \textsc{Singular} \cite{DGPS} with the workflow management system \textsc{GPI-Space} \cite{GSPC},
developed at the Fraunhofer Institute ITWM. It originates in an effort to realize massively parallel computations computer algebra, and has bee used for a variety of problems, for example, from commutative algebra, 
algebraic geometry, and tropical geometry \cite{singgpi,reinbold2018masters, boehm2020gitfan,bendle2020parallel,BFK21}.

In the present section, we discuss the application of the \textsc{Singular}/\textsc{GPI-Space} framework to the partial fraction decomposition. 
The application of the framework to 
the RREF problem has already been addressed in \cite{Bendle:2019csk}. Though our aim is an efficient parallel implementation of the partial fraction decomposition problem, the programming constructs developed along our way in terms of Petri nets are in fact useful in its own right and may be used in other algorithmic problems with a similar structure: While the \emph{WaitAll} workflow is obviously essential in countless settings where the problem splits into independent (possibly very unevenly sized) subproblems, the \emph{Fork/Join} workflow  will, for example, also be applicable to modular computations relying on a Chinese Remainder lift. 

\subsection {\textsc{GPI-Space and Singular}}
The task-based workflow management system \textsc{GPI-Space} is based on Gelernter's approach of separating coordination and computation \cite{gelernter}. The idea is that through a coordination language, a computation can convey information to another computation whose state is evolving and unpredictable. 

To realize an implementation of this approach, \textsc{GPI-Space} provides three main components: a distributed runtime system, that manages the resources (assigning jobs to cores); a virtual memory layer, which allows computations to access common data, and a workflow engine, which provides the coordination layer. In the coordination layer, the user specifies a program relying on the language of Petri nets~\cite{petrinet1962}:
\begin{defn}
	A \emph{Petri net} is a bipartite directed graph $N = (P,T,F)$, where
	where  $P $ and $T$ are disjoint, finite sets, and
	$$F \subseteq (P \times T) \cup (T \times P). $$
  The set $P$ contains the \emph{places} of the net,
	the set $T$ the \emph{transitions}. The set $F$ is called
	the \emph{flow relation} of the net. If $(p, t) \in F$ then we say $p$ is
	an \emph{input} of $t$ and if $(t, p) \in F$ we say $p$ is an \emph{output} of $t$.
\end{defn}

In addition to this static part of the concept of Petri nets, there are dynamic aspects describing the execution of the net.\begin{defn}
	A \emph{marking} of $N$ is a map $M:P \rightarrow \mathbb{N}_{0}$. 
	\end{defn}
A marking defines the state of the Petri net, and may be thought of as a map that counts a number of tokens on a
given place. 
\begin{defn}
	 We say that $M$ \emph{enables} the  
	transition $t$ of $N$ if, for all $p$ such that $(p,
	t) \in F$, we have $M(p) > 0$.
\end{defn}
If a Petri net equipped with a marking has an enabled transition,
the transition can be \emph{fired}. Firing is a process which maps a given marking
marking $M$  to a new marking $M'$, defined by

\[
	M'(p) := M(p) - \lvert \left\{ (p,t) \right\}  \cap F\rvert
	+ \lvert \left\{ (t,p) \right\}  \cap F\rvert.
\]
for all $p \in P$. Firing a transition can be thought of as that transition consuming a token from
each input and placing a new token on each output. A Petri net with a marking is \emph{executed} by firing random enabled transitions. 

In its implementation, \textsc{GPI-Space} expands the concept of a Petri net  in the sense that
tokens can be complex data structures. Transitions run code on
that data and the result usually determines the data carried by the tokens placed on the
output places of the respective transition. \textsc{GPI-Space} also allows the user to put conditionals on transitions: A
transition can inspect the contents of an input token without consuming it (yet)
and only accept it provided the stated conditions are met.


Fundamental functionality of our \textsc{Singular} library that implements the algorithmic building blocks of the partial fraction decomposition
is executed by the C-library version of \textsc{Singular} and is then wrapped
into transitions of a \textsc{GPI-Space} Petri net that models the coordination level structure of the algorithm.  Our implementation in \textsc{GPI-Space} is, in turn, configured and called from a standard \textsc{Singular} for transparent and convenient user interaction.

\subsection{Algorithmic realization of the partial fraction decomposition based on Petri nets}\label{Petrinet}

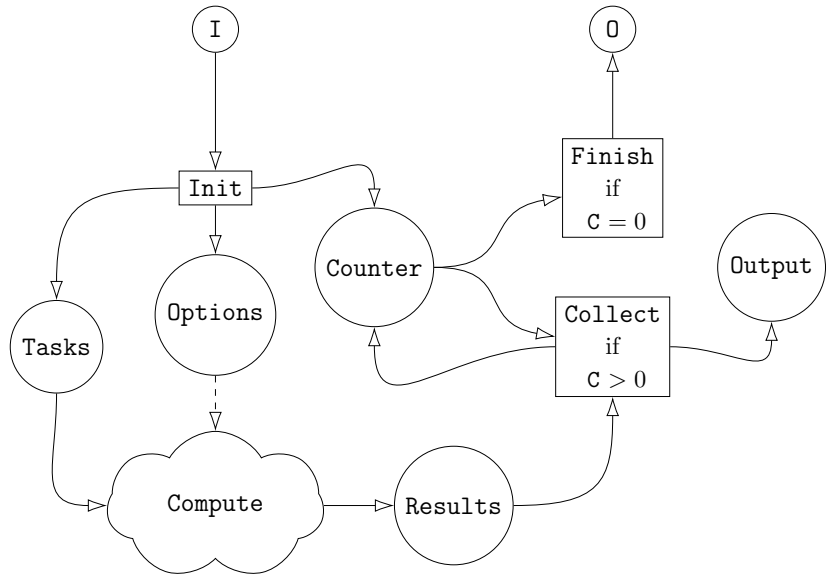
\begin{figure}[tbh]
	\centering
	         \resizebox{0.8\textwidth}{!}{
		\begin{tikzpicture}[
			place/.style = {draw, circle},
			trans/.style = {draw, rectangle},
			net/.style = {cloud, draw,cloud puffs=7,cloud puff arc=120, aspect=2, inner ysep=1em},
			arrow/.style = {-{Latex[open, length=3mm, width=2mm]}},
			every text node part/.style={align=center}
			]
		\def \u {2.5}
		\node[place] (I)       at ( 0 * \u,  2 * \u) {\texttt{I}};
		\node[place] (Options) at ( 0 * \u,  0.2 * \u) {\texttt{Options}};
		\node[trans] (Init)    at ( 0 * \u,  1 * \u) {\texttt{Init}};
		\node[place] (Tasks)   at (-1 * \u,  0 * \u) {\texttt{Tasks}};
		\node[net]   (Compute) at ( 0 * \u, -1 * \u) {\texttt{Compute}};
		\node[place] (Results) at (1.5* \u, -1 * \u) {\texttt{Results}};
		\node[place] (Count)   at ( 1 * \u,  0.5 * \u) {\texttt{Counter}};
    \node[trans] (Collect)  at ( 2.5 * \u,  0 * \u)
          {\texttt{Collect} \\
                   \small{if} \\
                   \small{ \texttt{C} $> 0$ }};
		\node[place] (Done)   at ( 3.5 * \u,  0.5 * \u) {\texttt{Output}};
    \node[trans] (Finish)  at ( 2.5 * \u,  1 * \u)
          {\texttt{Finish} \\
                   \small{if} \\
                   \small{ \texttt{C} $= 0$ }};
    \node[place] (Output)  at ( 2.5 * \u,  2 * \u) {\texttt{O}};

    \draw[arrow]
      (I)
      to
      (Init);
    \draw[arrow]
      (Init)
      to [out=180, in=90, looseness=1.5]
      (Tasks);
    \draw[arrow]
      (Init)
      to
      (Options);
    \draw[arrow]
      (Init)
      to [out=0, in=90, looseness=1.5]
      (Count);
    \draw[dashed,arrow]
      (Options)
      to
      (Compute);
    \draw[arrow]
      (Tasks)
      to [out=270, in=180, looseness=1.5]
      (Compute);
    \draw[arrow]
      (Compute)
      to
      (Results);
    \draw[arrow]
      (Results) to
      [out=0, in=270, looseness=1.5]
      (Collect);
    \draw[arrow]
      (Count)
      to [out=0, in=170, looseness=1.5]
      (Collect);
    \draw[arrow]
      (Collect)
      to [out=180, in=270, looseness=1.5]
      (Count);
    \draw[arrow]
      (Collect)
      to [out=0, in=270, looseness=1.5]
      (Done);
    \draw[arrow]
      (Count)
      to [out=0, in=190, looseness=1.5]
      (Finish);
    \draw[arrow]
      (Finish)
      to [out=90, in=270, looseness=1.5]
      (Output);

	\end{tikzpicture}}
	\caption{The \texttt{WaitAll} Petri net to parallelize over an input tuple}
	\label{fig:parallel_all}
\end{figure}
\begin{figure}[tbh]
	\centering
	 \resizebox{0.8\textwidth}{!}{
		\begin{tikzpicture}[
			place/.style = {draw, circle},
			trans/.style = {draw, rectangle},
			net/.style = {cloud, draw,cloud puffs=7,cloud puff arc=120, aspect=2, inner ysep=1em},
			arrow/.style = {-{Latex[open, length=3mm, width=2mm]}},
			every text node part/.style={align=center}
			]
		\def \u {2.5}
		\node[place] (I)          at ( -1 * \u,  3 * \u)  {\texttt{I}};
		\node[trans] (Gen_prep)   at (-1 * \u,  2 * \u)  {\texttt{Prepare}};
		\node[place] (Done)       at (-1 * \u,  1 * \u)  {\texttt{Tasks}};
    \node[trans] (Coll_done)  at ( 1 * \u,  1 * \u)  {\texttt{Collect} \\
                                                      \small{if} \\
                                                  \small{ \texttt{Tasks.status}}
                                                  \small{$= 1$ }};
		\node[trans] (Handover)    at (-1 * \u, 0 * \u)  {\texttt{HandToStart}\\
                                                  \small{if} \\
                                                  \small{ \texttt{Tasks.status}}
                                                  \small{$= 0$ }};
    \node[place] (Prepared1)  at (-1 * \u, -1 * \u)  {};
		\node[trans] (NSS)        at (-1 * \u, -2 * \u)  {\texttt{Nullstellensatz}};
		\node[place] (NSS_done)   at ( 1 * \u, -2 * \u)  {};
    \node[net]   (Short_num)  at ( 3 * \u, -2 * \u)  {\texttt{Short Numerator} \\
                                                      \texttt{Decomposition}};
		\node[place] (SN_done)    at ( 3 * \u, -1 * \u)  {};
    \node[net]   (Algebraic)  at ( 3 * \u,  0 * \u)  {\texttt{Algebraic} \\
                                                      \texttt{Dependence}};
		\node[place] (Alg_done)   at ( 3 * \u,  1 * \u)  {};
		\node[net]   (Numeric)    at ( 3 * \u,  2 * \u)  {\texttt{Numerator} \\
                                                      \texttt{Decomposition}};
		\node[place] (Num_done)   at ( 3 * \u,  3 * \u)  {};
		\node[trans] (Write)      at ( 2 * \u,  3 * \u)  {\texttt{Write}};
    \node[place] (Output)     at ( 1 * \u,  3 * \u) {\texttt{O}};


    \draw[arrow]
      (I)
      to
      (Gen_prep);

    \draw[arrow]
      (Gen_prep)
      to
      (Done);

    \draw[arrow]
      (Done)
      to
      (Coll_done);

    \draw[arrow]
      (Coll_done)
      to
      (Output);

    \draw[arrow]
      (Done)
      to
      (Handover);

    \draw[arrow]
      (Handover)
      to
      (Prepared1);

    \draw[arrow]
      (Prepared1)
      to
      (NSS);

    \draw[arrow]
      (NSS)
      to
      (NSS_done);

    \draw[arrow]
      (NSS_done)
      to
      (Short_num);

    \draw[arrow]
      (Short_num)
      to
      (SN_done);

    \draw[arrow]
      (SN_done)
      to
      (Algebraic);

    \draw[arrow]
      (Algebraic)
      to
      (Alg_done);

    \draw[arrow]
      (Alg_done)
      to
      (Numeric);

    \draw[arrow]
      (Numeric)
      to
      (Num_done);

    \draw[arrow]
      (Num_done)
      to
      (Write);

    \draw[arrow]
      (Write)
      to
      (Output);

	\end{tikzpicture}}
	\caption{The subnet calculating the partial fraction decomposition.}
	\label{fig:parallel_pfd}
\end{figure}
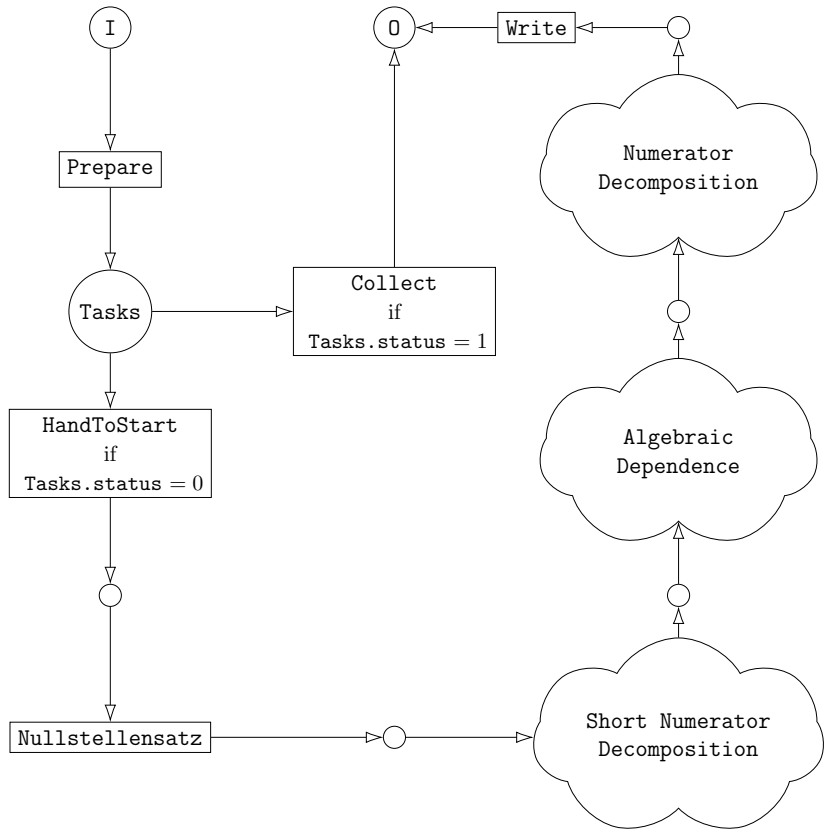
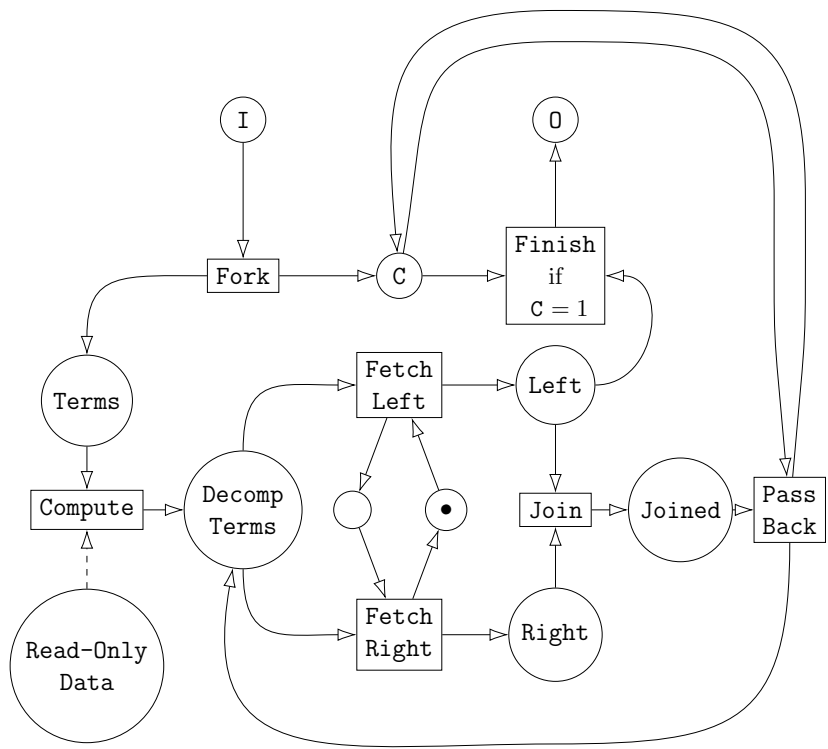
\begin{figure}[tbh]
	\centering
	 \resizebox{0.8\textwidth}{!}{
		\begin{tikzpicture}[
			place/.style = {draw, circle},
      empty/.style = {draw,circle, minimum size=0.6cm},
			trans/.style = {draw, rectangle},
			arrow/.style = {-{Latex[open, length=3mm, width=2mm]}},
			every text node part/.style={align=center}
			]
		\def \u {2.5}
		\node[place] (I)       at ( 0 * \u,  1.5 * \u) {\texttt{I}};
		\node[place] (Options) at ( -1 * \u,  -2 * \u) {\texttt{Read-Only} \\
                                                    \texttt{Data}};
		\node[trans] (Split)   at ( 0 * \u,  0.5 * \u) {\texttt{Fork}};
		\node[place] (Terms)   at (-1 * \u,  -0.3 * \u) {\texttt{Terms}};
		\node[trans] (Compute) at (-1 * \u, -1 * \u) {\texttt{Compute}};
    \node[place] (Decomp)  at ( 0 * \u, -1 * \u)  {\texttt{Decomp}\\
                                                  \texttt{Terms}};
    \node[trans] (Fetchl)  at ( 1 * \u,-0.2 * \u)  {\texttt{Fetch}\\
                                                  \texttt{Left}};
    \node[trans] (Fetchr)  at ( 1 * \u,-1.8 * \u)  {\texttt{Fetch}\\
                                                  \texttt{Right}};
    \node[place] (lactive) at (1.3 * \u,-1 * \u)  {\textbullet};
    \node[empty] (ractive) at (0.7 * \u,-1 * \u)  {};
    \node[place] (Left)    at ( 2 * \u,-0.2 * \u)  {\texttt{Left}};
    \node[place] (Right)   at ( 2 * \u,-1.8 * \u)  {\texttt{Right}};
    \node[trans] (Merge)   at ( 2 * \u, -1 * \u)
          {\texttt{Join}};
    \node[place] (Merged)   at ( 2.8 * \u,-1 * \u)  {\texttt{Joined}};
      \node[trans] (PassB)   at ( 3.5 * \u,-1 * \u)  {\texttt{Pass}\\
                                                      \texttt{Back}};
		\node[place] (Count)   at ( 1 * \u,  0.5 * \u) {\texttt{C}};
    \node[trans] (Finish)  at ( 2 * \u,  0.5 * \u)
          {\texttt{Finish} \\
                   \small{if} \\
                   \small{ \texttt{C} $= 1$ }};
    \node[place] (Output)  at ( 2 * \u,  1.5 * \u) {\texttt{O}};

    \draw[arrow]
      (I)
      to
      (Split);
    \draw[arrow]
      (Split)
      to [out=180, in=90, looseness=1.5]
      (Terms);
    \draw[arrow]
      (Split)
      to
      (Count);
    \draw[dashed,arrow]
      (Options)
      to
      (Compute);
    \draw[arrow]
      (Terms)
      to
      (Compute);
    \draw[arrow]
      (Compute)
      to [out=0, in=180, looseness=1.5]
      (Decomp);
    \draw[arrow]
      (Count)
      to [out=0, in=180, looseness=1.5]
      (Finish);
    \draw[arrow]
      (Count)
      to [out=80, in=180, looseness=1.5]
      (2 * \u, 2.0 * \u)
      to [out=0, in=90, looseness=1.5]
      (3.4 * \u, 1 * \u)
      to [out=270, in=95, looseness=1.5]
      (PassB);
    \draw[arrow]
      (PassB)
      to [out=85, in=270, looseness=1.5]
      (3.6 * \u, 1 * \u)
      to [out=90, in=0, looseness=1.5]
      (2 * \u, 2.2 * \u)
      to [out=180, in=95, looseness=1.5]
      (Count);
    \draw[arrow]
      (Decomp)
      to [out=90, in=180, looseness=1.5]
      (Fetchl);
    \draw[arrow]
      (Decomp)
      to [out=270, in=180, looseness=1.5]
      (Fetchr);
    \draw[arrow]
      (lactive)
      to
      (Fetchl);
    \draw[arrow]
      (Fetchl)
      to
      (ractive);
    \draw[arrow]
      (ractive)
      to
      (Fetchr);
    \draw[arrow]
      (Fetchr)
      to
      (lactive);
    \draw[arrow]
      (Fetchl)
      to
      (Left);
    \draw[arrow]
      (Fetchr)
      to
      (Right);
    \draw[arrow]
      (Left)
      to
      (Merge);
    \draw[arrow]
      (Right)
      to
      (Merge);
    \draw[arrow]
      (Merge)
      to
      (Merged);
    \draw[arrow]
      (Merged)
      to
      (PassB);
    \draw[arrow]
      (PassB)
      to [out=270, in=0, looseness=1.5]
      (2 * \u , -2.5 * \u)
      to [out=180, in=260 ,looseness=1.5]
      (Decomp);
    \draw[arrow]
      (Left)
      to [out=0, in=0, looseness=1.5]
      (Finish);
    \draw[arrow]
      (Finish)
      to [out=90, in=270, looseness=1.5]
      (Output);

	\end{tikzpicture}}
	\caption{The \texttt{Fork/Join} subnet to parallelize over the decomposition of
  terms.}
	\label{fig:parallel_merge}
\end{figure}

The elements of a set of rational function coefficients are decomposed using a workflow implemented in \textsc{GPI-Space}. 
Parallelism is achieved by two mechanisms: first, the Petri net
processes different functions in parallel using a \texttt{WaitAll}
approach, which is depicted in Figure \ref{fig:parallel_all}, and second, by
parallelizing the algorithmic sub-steps of the PFD algorithm applied to an individual function using a
\texttt{Fork/Join} approach on the terms, which is implemented in the Petri nets shown in Figures
 \ref{fig:parallel_pfd} and \ref{fig:parallel_merge}.
Note that the concept a of Petri net is modular, and the Petri net in Figure \ref{fig:parallel_merge} is a subnet of that depicted
in Figure \ref{fig:parallel_pfd}, taking the place of the transition/subnet depicted
as a cloud. The net in Figure \ref{fig:parallel_pfd} 
is in turn a subnet of that
in Figure~\ref{fig:parallel_all}. The overall Petri net leads to an intertwined form of parallelism which is able to handle the decomposition of terms of different IBP coefficients at different stages of the PFD algorithm in parallel, which leads to an efficient utilization of the resources und thus a good parallel efficiency.

We remark that Figures \ref{fig:parallel_all} and \ref{fig:parallel_merge} have a
similar structure, with the major addition of a parallel merge mechanism in the
latter. In Figure \ref{fig:parallel_merge},  the results of the parallel computations are all part of the
same problem and hence need to be recombined, whereas in Figure \ref{fig:parallel_all}, the
results are
unrelated and treated individually.

In the following, we briefly discuss each Petri net by listing first the
places, followed by  the transitions that connect to the various places. 

We begin
with Figure~\ref{fig:parallel_all}.
	Place \texttt{I} is initialized with an input token provided by the client
  program (\textsc{Singular} implementing the user interface), and contains, in
  particular, the given set of rational functions.  The transition \texttt{Init} then
  uses this data to set up various other places: The tokens on \texttt{Tasks} 
  each represent a function to be decomposed. 
  The single token on the
  place~\texttt{Counter} has a field storing the total number of
  computations to be performed and a field counting the number of finished
  computations which is initialized with zero. The place \texttt{Options} holds
  a token with read-only input data and auxiliary technical data for the compute transition.\footnote{Note that, if we want to copy data from an input token of a transition to an output token, it is possible and typically preferable to only copy a reference to the actual data.} Tokens on the
  place \texttt{Results} will represent finished computations.
	%
	%
	%
    %
  These finished computations will be passed to the place \texttt{Output},
  where the client program can then collect them.
	The place \texttt{O} will be used
		to indicate by the presence of a token that the Petri net has finished with
    all computations ready for collection at \texttt{Output}.\footnote{Data in the token may be used to pass, for example, debugging information
    to \textsc{Singular}.}
  In our application, the transition/subnet 
  \texttt{Compute}
    calls the Petri net depicted in Figure~\ref{fig:parallel_pfd}.  
    The transition/subnet    consumes a token from 
    \texttt{Tasks} and reads the values at
		\texttt{Options}.  This information is then used to
		perform the desired partial fraction decomposition for the
		respective function. 
   Once a
    computation has finished, a token is placed on \texttt{Results}.
  The \texttt{Collect} transition fires conditionally on the token on
    \texttt{Counter} being strictly greater than $0$, that is, that not all
    computations that were started have finished. In that case, it
    consumes a token from \texttt{Results}, decrements the token on \texttt{Counter} by
    one, and passes the token on to the place \texttt{Output}.
  Once the value of \texttt{C} equals $0$, the
    condition for the transition \texttt{Finish} is met, and the transition
    places a token on \texttt{O}. The client program then collects the results
    on the place \texttt{Output}.


  The net in Figure \ref{fig:parallel_pfd} is a subnet of Figure
    \ref{fig:parallel_all} representing \texttt{Compute} and, hence, must take as
    input and output the same token types as the \texttt{Compute} transition/subnet in Figure
    \ref{fig:parallel_all}.
The place \texttt{I} in Figure~\ref{fig:parallel_pfd} takes rational functions.
    Although not displayed explicitly, information on the place \texttt{Options}
    of Figure  \ref{fig:parallel_all} is used by transitions in Figure
    \ref{fig:parallel_pfd}.
  Tokens on the place \texttt{Tasks} hold two fields, one corresponding to
    the rational function and a boolean flag \texttt{status},
    indicating that no further computation on this function is needed
   . 
  The places between the computational steps are unmarked, as these only hold the tokens passed on between the steps.
	Once all the steps are completed for a given function, the corresponding result is
    placed on \texttt{O}.

  The transition \texttt{Prepare} consumes the input token provided on \texttt{I} and
    prepares the input data so that we can work in the decomposition steps with a consistent input/output format.  As part of this process, it checks whether the
    rational function is zero or has
    constant denominator, in which case the  the data field \texttt{status} of the token put on \texttt{Tasks} is given
    the value $1$  to
    indicate that no further computation is needed. In this case, the
    \texttt{Collect} transition consumes the token at \texttt{Tasks} and passes it on to \texttt{O}. Otherwise, \texttt{Prepare} assigns the
    \texttt{status} value
    $0$ to the token (indicating the computation is not yet complete), which is then handed on to the input place for the transition
    \texttt{Nullstellensatz}. This transition computes the
    Nullstellensatz decomposition in \textsc{Singular} by applying the respective
    function provided in our library \texttt{pfd.lib}.

  The Petri net depicted in Figure \ref{fig:parallel_merge} is used to realize each of the decomposition steps \texttt{ShortNumeratorDecomposition},
    \texttt{AlgebraicDependence} and \texttt{NumeratorDecomposition} in a parallel way over the
    summands received as input, relying on a \texttt{Fork/Join} approach:  The
    list of
    input summands is split into individual tokens, which are then decomposed
    in parallel. The resulting
    lists of summands are merged before being handed over to the
    next transition. The merge step at the end of such a decomposition often leads
    to cancelation, which reduces the size of the output and speeds up the
    computation. Note that, as the fork process is done in our approach with respect to the summands of the input, this does not lead to additional parallelism for the \texttt{Nullstellensatz} transition, that is, this transition is parallel only over the different rational functions.
Finally, the \texttt{Write} transition writes the resulting decomposition in
    various serialization formats, before handing a token to \texttt{O} to signal that the algorithm has finished.

We discuss the \texttt{Fork/Join} process  illustrated in Figure~\ref{fig:parallel_merge} in more detail.  The aim is to fork into a number of
    parallel computations, and  join two results into one via a parallel approach until only a single token is left, which then is put on the
    output place. 
   The \texttt{Fork} step takes a
    token representing a list of terms (corresponding to a sum) and splits it into a partition of sublists of approximately equal
    length, which in turn are handled in parallel. 
Once a \texttt{Compute} task has been finished, the corresponding token with a list of terms is
  put on
  \texttt{DecompTerms}. A pair of transitions - \texttt{FetchLeft} and
  \texttt{Fetch Right} - are responsible for dividing these summands evenly into
    the two input places for the \texttt{Join} transition, which combines
    the two lists into one (with possibly cancellation occurring) and puts the resulting token on its output place \texttt{Joined}. The place
    \texttt{C} is used to keep count of the number of tokens still to be merged
    and to enable the \texttt{Finish} transition once there is only a single
    token remaining.
 Consuming a token from \texttt{Joined}, the transition \texttt{PassBack}
    decrements the counter for the number of parallel results to still to be
    merged by one and places the token onto \texttt{DecompTerms}. 
    
    The fetch
    transitions ensure that whenever a token is put on \texttt{Left}, the
    following token is placed  on \texttt{Right} by passing a control token
    between two control places. A trivial inductive argument on the number of tokens on
    \texttt{Left}, \texttt{Right} and \texttt{DecompTerms} shows that once all
    the summands have been merged together, the final token will be placed at
    \texttt{Left}. 
    At this point, the \texttt{Finish} step will collect the
    token and place a finishing token at \texttt{O}. 
        
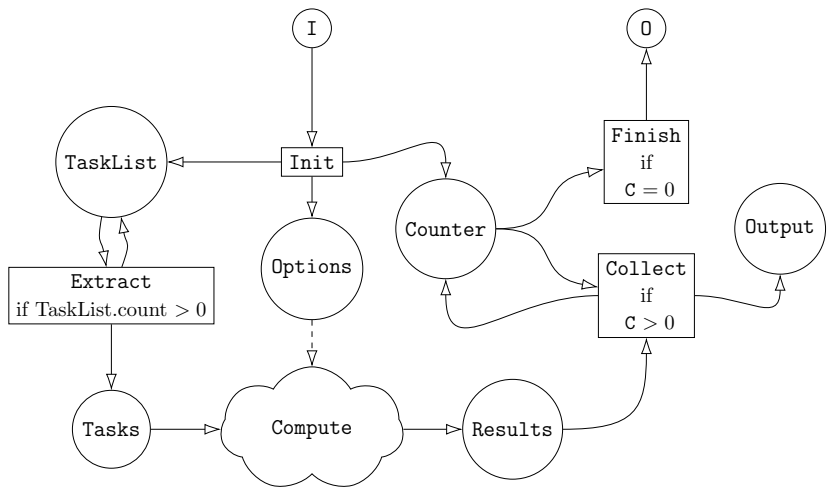
\begin{figure}
	\centering
	\resizebox{0.8\textwidth}{!}{
		\begin{tikzpicture}[
			place/.style = {draw, circle},
			trans/.style = {draw, rectangle},
			net/.style = {cloud, draw,cloud puffs=7,cloud puff arc=120, aspect=2, inner ysep=1em},
			arrow/.style = {-{Latex[open, length=3mm, width=2mm]}},
			every text node part/.style={align=center}
			]
		\def \u {2.5}
		\node[place] (I)       at ( 0 * \u,  2 * \u) {\texttt{I}};
		\node[place] (Options) at ( 0 * \u,  0.2 * \u) {\texttt{Options}};
		\node[trans] (Init)    at ( 0 * \u,  1 * \u) {\texttt{Init}};
		\node[place] (TaskList) at (-1.5 * \u,  1 * \u) {\texttt{TaskList}};
		\node[trans] (Extract)  at (-1.5 * \u,  0 * \u) {\texttt{Extract} \\
                                                    \small{if}
                                                    \small{TaskList.count $> 0$}};
		\node[place] (Tasks)   at (-1.5 * \u,  -1 * \u) {\texttt{Tasks}};
		\node[net]   (Compute) at ( 0 * \u, -1 * \u) {\texttt{Compute}};
		\node[place] (Results) at (1.5* \u, -1 * \u) {\texttt{Results}};
		\node[place] (Count)   at ( 1 * \u,  0.5 * \u) {\texttt{Counter}};
    \node[trans] (Collect)  at ( 2.5 * \u,  0 * \u)
          {\texttt{Collect} \\
                   \small{if} \\
                   \small{ \texttt{C} $> 0$ }};
		\node[place] (Done)   at ( 3.5 * \u,  0.5 * \u) {\texttt{Output}};
    \node[trans] (Finish)  at ( 2.5 * \u,  1 * \u)
          {\texttt{Finish} \\
                   \small{if} \\
                   \small{ \texttt{C} $= 0$ }};
    \node[place] (Output)  at ( 2.5 * \u,  2 * \u) {\texttt{O}};

    \draw[arrow]
      (I)
      to
      (Init);
    \draw[arrow]
      (Init)
      to [out=180, in=0, looseness=1.5]
      (TaskList);
    \draw[arrow]
      (TaskList)
      to [out=260, in=100, looseness=1.5]
      (Extract);
    \draw[arrow]
      (Extract)
      to [out=70, in=280, looseness=1.5]
      (TaskList);
    \draw[arrow]
      (Extract)
      to
      (Tasks);
    \draw[arrow]
      (Init)
      to
      (Options);
    \draw[arrow]
      (Init)
      to [out=0, in=90, looseness=1.5]
      (Count);
    \draw[dashed,arrow]
      (Options)
      to
      (Compute);
    \draw[arrow]
      (Tasks)
      to
      (Compute);
    \draw[arrow]
      (Compute)
      to
      (Results);
    \draw[arrow]
      (Results) to
      [out=0, in=270, looseness=1.5]
      (Collect);
    \draw[arrow]
      (Count)
      to [out=0, in=170, looseness=1.5]
      (Collect);
    \draw[arrow]
      (Collect)
      to [out=180, in=270, looseness=1.5]
      (Count);
    \draw[arrow]
      (Collect)
      to [out=0, in=270, looseness=1.5]
      (Done);
    \draw[arrow]
      (Count)
      to [out=0, in=190, looseness=1.5]
      (Finish);
    \draw[arrow]
      (Finish)
      to [out=90, in=270, looseness=1.5]
      (Output);

	\end{tikzpicture}}
	\caption{The \texttt{WaitAll} Petri net with pre-sorting by input file size}
	\label{fig sorting}
\end{figure}

    \begin{rem} If the input sizes of the rational functions are varying significantly, the randomized nature of the execution of the Petri net will lead to some drop in efficiency due to late scheduling of long-running jobs. This can be remedied by scheduling the hardest problems first. In Figure \ref{fig sorting} we provide a variation of the \texttt{WaitAll} net that provides a simple but useful realization of this strategy. The transition \texttt{Init} generates a token with a list with the rational functions on the place \texttt{TaskList} such that the entries are sorted by byte size.
  The transition \texttt{Extract} extracts and deletes the entry of
  largest byte size, places it on \texttt{Tasks} and decrements 
  \texttt{TaskList.count} by one.
 \end{rem}

\section{Instructions on how to use the massively parallel PFD application}
\label{sec:pfd_usage}
Our framework can be used from the interactive \textsc{Singular} command line or from within \textsc{Singular} libraries. In order to do so, the \textsc{pfd-parallel} package and its dependencies have to be installed. This includes   \textsc{GPI-Space}, \textsc{GPI-2},  and \textsc{Singular}. 

We provide a one-line-installation for Linux environments based on the \textsc{Spack} package manager \cite{spack}. 
Detailed instructions can be found in the Github repository of our
application:

\begin{center}
\href{https://github.com/singular-gpispace/pfd-parallel}{https://github.com/singular-gpispace/pfd-parallel}
\end{center}
These instructions are self contained in the sense that they include the installation of \textsc{Spack}, and the installation of our package wkth installation of all dependencies. The project has been tested on Centos $7$ and $8$ and Ubuntu $18.04$ LTS and $20.04$ LTS. While we also provide manual installation instructions, we recommend the use of Spack, since it takes care of all dependencies automatically.

Assume that \texttt{\$\{example\_ROOT\}} has been
set to some directory accessible from all nodes involved in the installation and computation.
The Spack install will set, relying on \texttt{\$\{example\_ROOT\}}, the following environment variables: \texttt{\$PFD\_INSTALL\_DIR}, which is the directory where our
application was installed, and \texttt{\$SINGULAR\_INSTALL\_DIR}, which is the directory
where \textsc{Singular} was installed.
Change directory to \texttt{\$\{example\_ROOT\}}:
\begin{lstlisting}[language=bash]
cd ${example_ROOT}
\end{lstlisting}
To run our example, we have to create some configuration and input data in \texttt{\$\{example\_ROOT\}}:
\begin{itemize}[leftmargin=*]
  \item
    A nodefile with the machines to be used as nodes (one line per node). For example, the following
    command generates such a file with just the current machine:
\begin{lstlisting}[language=bash]
hostname > hostfile
\end{lstlisting}
  \item
    Optionally, the \textsc{GPI-Space} Monitor can be started to display computations in form of a Gantt diagram (to use the monitor, you need an X-Server running).
    In case you do not want to use the monitor, you should not set in \textsc{Singular}
    the fields \texttt{options.loghostfile} and \texttt{options.logport} of the
    \textsc{GPI-Space} configuration token (see below). In order to use the
    \textsc{GPI-Space} monitor, we need a \texttt{loghostfile} with the address of the
    machine running the monitor.
\begin{lstlisting}[language=bash]
hostname > loghostfile
\end{lstlisting}
    On this machine, we start the monitor, specifying a TCP port where the
    monitor will receive information from \textsc{GPI-Space}. The same port has
    to be specified in the configuration token in the field \texttt{options.logport} (see below).
\begin{lstlisting}[language=bash]
${PFD_INSTALL_DIR}/libexec/bundle/gpispace/bin/gspc-monitor --port 9876
\end{lstlisting}
  \item
    To create some example files with input, we copy data included in the PFD install directory: 
\begin{lstlisting}[language=bash]
mkdir -p ${example_ROOT}/input
mkdir -p ${example_ROOT}/results
cp -v ${PFD_INSTALL_DIR}/example_data/* ${example_ROOT}/input
\end{lstlisting}
    These files contain the entries of a 1x10 row of  matrix of rational functions
    for which the partial fraction decomposition should be computed.
  \item
    Finally, a directory for temporary files is need. This directory will be used during runtime, and should be
    accessible from all machines involved in the computation:
\begin{lstlisting}[language=bash]
mkdir ${example_ROOT}/tempdir
\end{lstlisting}

\end{itemize}
We can now start \textsc{Singular}, telling it where to find the library and the shared object file for
the PFD application:
\begin{lstlisting}[language=bash]
SINGULARPATH="$PFD_INSTALL_DIR/LIB"  $SINGULAR_INSTALL_DIR/bin/Singular
\end{lstlisting}
In the \textsc{Singular} interpreter shell, enter the following code to setup the framework (or make similar code part of a library you are writing):
\begin{lstlisting}[language=C]
LIB "pfd_gspc.lib";
configToken gspcconfig = configure_gspc();
gspcconfig.options.tempdir = "tempdir";
gspcconfig.options.nodefile = "hostfile";
gspcconfig.options.procspernode = 8;
\end{lstlisting}
If the \textsc{GPI-Space} monitor is supposed to be used then also do:
\begin{lstlisting}[language=C]
gspcconfig.options.loghostfile = "loghostfile";
gspcconfig.options.logport = 9876;
\end{lstlisting}
We also define a configuration token for the PFD algorithm:
\begin{lstlisting}[language=C]
configToken pfdconfig = configure_pfd();
pfdconfig.options.inputdir = "input";
pfdconfig.options.filename = "xb_deg5";
pfdconfig.options.suffix = "ssi";
pfdconfig.options.parallelism = "intertwined";
pfdconfig.options.algorithm = "Leinartas";
pfdconfig.options.outputformat = "ssi,cleartext,indexed_numerator_denominator";
pfdconfig.options.outputdir = "results";
\end{lstlisting}
This will
\begin{itemize}[leftmargin=*]
  \item
    load the \textsc{Singular} library providing access to the \textsc{pfd-parallel} application,
  \item
    add information where to store temporary data (in the corresponding field \texttt{options.tempdir} of the \textsc{GPI-Space} configuration token) with the path corresponding to the directory created above,
  \item
    specify where to find the nodefile (in the field
    \texttt{options.nodefile} of the \textsc{GPI-Space} configuration token), and
  \item
    set how many processes per node\footnote{A version of the framework which allows the user to set the number of cores individually for each node is under develpment.} should be started (in the corresponding field
    \texttt{options.procspernode}  of the \textsc{GPI-Space} configuration token, usually one process per core, not taking
    hyper-threading into account; you may have to adjust according to your
    available resources),
  \item
    specify the address of the \textsc{GPI-Space} monitor (this step is optional),
   \item
    specify a base file name used for accessing the array of input files by appending the indices, along with the path where the input files are located, and the path where the output files should be written (in the respective fields \texttt{options.filename}, \texttt{options.inputdir}, and \texttt{options.outputdir} of the PFD configuration token)\footnote{Note that in more complex environments like a cluster, it is usually preferable to specify
absolute paths, in particular, for the nodefile and the temporary directory.
Specifying the results directory is optional, and only necessary if the result
directory should be different from the input directory.},
  \item specify the input format (field \texttt{options.inputformat}), see Remark \ref{rmk format} below for a discussion of the formats\footnote{Note that if \texttt{ssi} is specified the program assumes that the data is available in the high-performance \textsc{Singular}  serialization format ssi, while if \texttt{txt} is specified, and the respective ssi file is not present in the input folder, the program automatically converts the text format to ssi, and continues with the ssi format.},
  \item specify the output formats (field \texttt{options.outputformat}, see again Remark \ref{rmk format} below; note that multiple output formats can be requested at the same time),
  \item
    specify the parallelization strategy to be used (by specifying the field \texttt{options.parallelism}  of the PFD configuration token), where the field can take the values \texttt{waitAll}, \texttt{intertwined}, or \texttt{size_strategy} for a parallelization over the different rational functions, for using in addition parallelism for the individual rational functions, and for handling the rational functions serially or in parallel depending on the size of the input  coefficient, respectively;  in the latter case, \texttt{options.percentage} has to be set to an integer $0\leq p \leq 100$ to specify that the $p/100$ largest of all input functions should be processed with parallelism per individual function, and
 \item choose the algorithm to be used (\texttt{Leinartas} or \texttt{MultivariateApart} for our  \textsc{Singular} implementations of the respective algorithms) by specifying the field \texttt{options.algorithm}  of the PFD configuration token; note that this option affects only the sequential processing of  individual functions, while parallel processing of individual functions is always done with our parallel Leinartas algorithm (since the MultivariateApart algorithm is not parallel).
\end{itemize}

\begin{rem}\label{rmk format}
Note that the input can be specified in one of two formats: a human-readable
text-based format, or a data structure encoded in the high-performance \textsc{Singular} ssi serialization format.\footnote{While the ssi format contains the information about the polynomial ring containing the numerators and denominators, using the text format requires that in the controlling \textsc{Singular} instance the polynomial ring is already defined} 
Our implementation outputs results in a number of formats: the same ssi
based format as used for
input; a self contained, human-readable text-based mathematical expression
(\texttt{cleartext}); a human-readable, indexed, text-based mathematical expression, where the numerators and denominators are factorized
and the unique factors represented by indexed variables (\texttt{indexed\_numerator_denominator}); and a
similar format, where only the denominators are factorized and indexed (\texttt{indexed\_denominator}).
The correspondence between the indexed variables and the factors is encoded in
an additional file, so the output in the indexed formats consist of two files.
When analyzing the byte size of the results of our algorithms, we usually consider the sum of the size of both files
in the \texttt{indexed\_numerator_denominator} format.

\end{rem}
Once the framework and the algorithm are configured, we can run the computation, specifying the indices of the array of input files (generated above) which should be
    processed:
\begin{lstlisting}[language=C]
list listofentries = list( list(1,1), list(1,2), list(1,3), list(1,4), list(1,9), list(1,10) );
parallel_pfd( listofentries, gspcconfig, pfdconfig);
\end{lstlisting}

\section{Examples for partial fraction via large scale parallelizations}
\label{sec:examples}
In this section, we apply our multivariate
partial fractioning framework in the case two examples, simplifying rational functions occurring in IBP reduction coefficients and in analytic multiloop scattering amplitudes. We provide timings and an analysis of compression ratios to
demonstrate the power of our framework.

\subsection{Simplifying IBP reduction coefficients}

\begin{figure}[ht]
\centering
\includegraphics[width=0.5\textwidth]{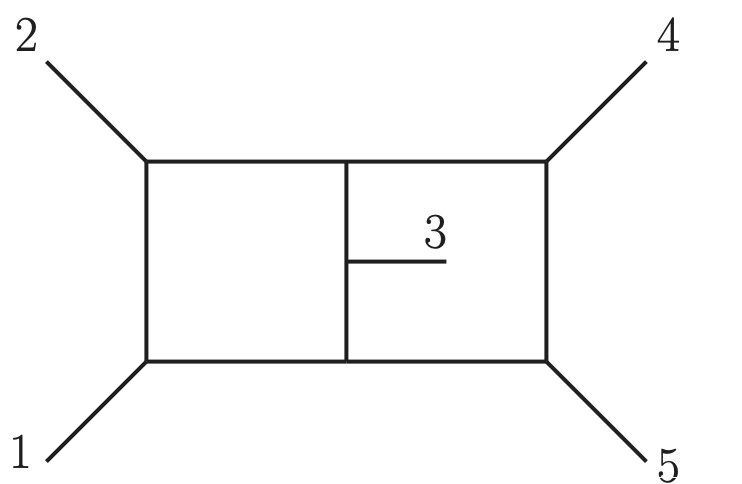}
\caption{Two-loop five-point nonplanar ``double pentagon'' diagram}
\label{double pentagon}
\end{figure}

We consider in this section the nonplanar two-loop five-point diagram
(Figure \ref{double pentagon}) as an example.
All external and internal lines are massless. The kinetic conditions are $2p_1\cdot p_2=s_{12}$, $2p_2\cdot p_3=s_{23}$, $2p_3\cdot p_4=s_{34}$, $2p_4\cdot p_5=s_{45}$ and $2p_1\cdot p_5=s_{15}$. The propagators are
\begin{equation}
\begin{aligned}
&D_1=l_1^2,\quad
D_2=(l_1-p_1)^2,\quad
D_3=(l_1-p_{12})^2,\quad
D_4=l_2^2,\\
&D_5=(l_2-p_{123})^2,\quad
D_6=(l_2-p_{1234})^2,\quad
D_7=(l_1-l_2)^2,\\
&D_8=(l_1-l_2+p_3)^2,\quad
D_9=(l_1-p_{1234})^2,\quad
D_{10}=(l_2-p_1)^2,\\
&D_{11}=(l_2-p_{12})^2,
\end{aligned}
\end{equation}
where $p_{i\dots j}=\sum_{k=i}^j p_k$. For this diagram, the UT basis
and the corresponding symbol form was given in refs.~\cite{Abreu:2018aqd,
Chicherin:2018old}, while
refs.~\cite{Chicherin:2018old,Chicherin:2020oor}  provided the
analytic evaluations. In ref.~\cite{Bendle:2019csk}, the analytic IBP reduction coefficients for
the integrals with ISP up to the degree $4$ in the top sector
were calculated. In ref.~\cite{Boehm:2020ijp}, using the
improved Leinartas' algorithm in Section
\ref{sec:algorithm}, the size of the IBP coefficients up to degree $4$ with respect to the UT basis was
decreased from 700MB to 19MB by rewriting the rational functions in terms of a partial fraction decomposition. Recently, integrals for this diagram with irreducible scalar product (ISP) of degree $5$ have been reduced to a well-chosen
basis in ref.~\cite{Klappert:2020nbg}, making use of the systems of
block triangular system
given in refs.~\cite{Liu:2017jxz,Liu:2018dmc,Guan:2019bcx}. In the choice of
basis of ref.~\cite{Klappert:2020nbg}, the analytic reduction
coefficients have a size of $\sim 25$ GB.

In this section, we reduce the relevant target integrals, that is, the $47$ top-sector integrals with ISPs up to degree $5$,
to a UT basis, and show that the coefficients can be greatly
simplified further by relying on our {\sc pfd-parallel} package. 

We use the module intersection method \cite{Bendle:2019csk} to obtain
a simple IBP system for reducing
target integrals. As has been already observed for the relations up to degree four, the number of IBP relations and corresponding size is much smaller than those generated by the
traditional Laporta algorithm. As shown with some more detail in Table
\ref{cut_IBP}, the total size of the system is only $17$ MB. The IBP generation step finishes within one hour on a single core. See also ref.~\cite{Guan:2019bcx} for
generating a simple integral relation system for the same diagram.
\begin{table}[ht] 
  \centering
  \caption{The IBPs system for the ``double-pentagon''
    integrals with numerator degree up to $5$, generated on the spanning cuts, relying on the module
    intersection method.}
  \label{cut_IBP}
  \ra{1.3}
  \medskip
  \begin{tabular}{@{}cccc@{}}
    \toprule
     Cut & \# relations & \# integrals & size\\
     \midrule

\{1,5,7\} &2723 & 2749& 1.4 MB \\
\{1,5,8\} &2753 & 2777& 1.6 MB  \\
\{1,6,8\} &2817 & 2822& 2.1 MB  \\
\{2,4,8\} &2918 & 2921& 2.1 MB  \\
\{2,5,7\} &2796 & 2805& 1.5 MB  \\
\{2,6,7\} &2769 & 2814 & 1.2 MB  \\
\{2,6,8\} &2801 & 2821 & 1.6 MB  \\
\{3,4,7\} &2742 & 2771 & 1.4 MB  \\
\{3,4,8\} &2824 & 2849 & 1.9 MB  \\
\{3,6,7\} &2662 & 2674 & 1.5 MB  \\
\{1,3,4,5\} & 1600 & 1650 & 0.72 MB \\
    \bottomrule
  \end{tabular}
\end{table}

We solve the IBP system generated via the module intersection method by numeric linear algebra and interpolation. Since the
IBP reduction coefficients tend to have simple denominators in a UT
basis~\cite{Boehm:2020ijp}, we interpolate the 
coefficients for a UT basis. The coefficients are rational
functions in $s_{12}, s_{23}, s_{23}, s_{23}, s_{23}, \epsilon$ and
$\epsilon_5$, where $\epsilon_5 = 4 \sqrt{-1} \epsilon_{\mu_1 \mu_2 \mu_3
  \mu_4} p_1^{\mu_1} p_2^{\mu_2}p_3^{\mu_3}p_4^{\mu_4}$ is from the UT
integral definition. 

As illustrated in \cite{Boehm:2020ijp}, the denominators of the IBP
reduction coefficients for this diagram when using a UT basis contain
factors either in $\epsilon$ or symbol letters. With a
semi-numeric computation, we determine all
irreducible factors in the denominators of the IBP
reduction coefficients and their expression in terms of the factors. For the example under consideration, the factors are:
\begin{equation}\label{dpUTpole}
\begin{aligned}
&\epsilon -1,\hspace{0.9em}
2 \epsilon -1,\hspace{0.9em}
3\epsilon -1,\hspace{0.9em}
4 \epsilon -1,\hspace{0.9em}
4\epsilon+1,\hspace{0.9em}
s_{12},\hspace{0.9em}
s_{15},\hspace{0.9em}
s_{15}-s_{23},\hspace{0.9em}
s_{23},\hspace{0.9em}\\
&s_{12}+s_{23},\hspace{0.9em}
s_{12}-s_{34},\hspace{0.9em}
s_{12}+s_{15}-s_{34},\hspace{0.9em}
s_{15}-s_{23}-s_{34},\hspace{0.9em}
s_{34},\hspace{0.9em} s_{23}+s_{34}\\
&s_{12}-s_{45},\hspace{0.7em}
s_{23}-s_{45},\hspace{0.7em}
s_{12}+s_{23}-s_{45},\hspace{0.7em}
s_{12}-s_{15}+s_{23}-s_{45}, \hspace{0.7em}
s_{12}-s_{34}-s_{45}, \\
&s_{12}+s_{15}-s_{34}-s_{45},\hspace{0.9em}
s_{45},\hspace{0.9em}
s_{15}-s_{23}+s_{45},\hspace{0.9em}
s_{34}+s_{45}, \hspace{0.9em} \epsilon_5
\end{aligned}
\end{equation}
Knowing the denominators, we analytically calculate the numerators
via polynomial interpolation  (which is computationally
easier than rational function interpolation). To perform the computation, we
take $s_{12}=1$, and determine the maximum degrees in the other variables
by numertic tests.  We then do the reduction with symbolic variables $s_{45}$
and $\epsilon$, and $307200$ numerical tuples of values for the remaining
variables. Via polynomial interpolation, the analytic expressions of the IBP reduction
coefficients are obtained. The variable $s_{12}$ is then restored. The interpolation of the numerators finishes within $3$ hours on $210$
CPU cores. Restoring $s_{12}$ by dimensional analysis can be done within $3$ hours on $20$
CPU cores.  The resulting analytic expressions for the reduction coefficients form a
$47\times 108$ matrix, which takes about $19.6$ GB of disk space.

\begin{figure}[h]
    \caption{A scatterplot of the run times of the improved Leinartas algorithm against the input file sizes for the IBP reduction coefficients for the five-point non-planar ``double pentagon'' diagram for the degree up to $5$.}
    \label{fig:xbdeg_pfd_scatter_size_vs_runtime}
    \centerline{\includegraphics[height=8cm]{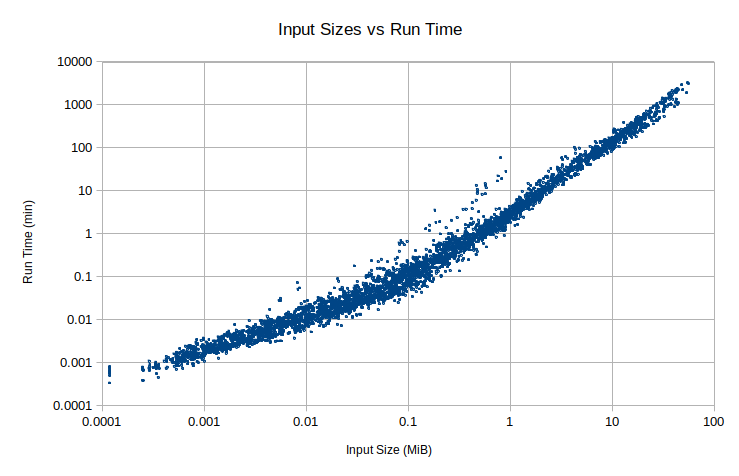}}
    \centering
  \end{figure}

We then use {\sc pfd-parallel} to simplify this huge set of
coefficients. When using one layer of
parallelization over the different coefficients, our implementation of the improved Leinartas algorithm can simplify the $19.6$
GB data set of the original reduction coefficients to expressions of a total size of $186$ MB within $68$
hours of parallel computation on $350$ cores.\footnote{We use a cluster consisting of 10 nodes with two Intel Xeon Gold 6230R CPUs and 768 GB RAM each.} We remark that using the improved Leinartas algorithm or the \textsc{MultivariateApart} algorithm (both implemented in our framework and available as computational backend) produce a similar  output size and computation time. Figure~\ref{fig:xbdeg_pfd_scatter_size_vs_runtime} shows a scatterplot of the run times of the improved Leinartas algorithm against the input file sizes for the IBP reduction coefficients

  While this form of parallelism of course provides a significant speedup compared to a sequential computation as well as compared to a parallel computation relying on an a priori partitioning of the set of jobs into batches,
  the approach is not as efficient as one would desire: Most of the coefficients require only less
  then $3$ hours to finish, and our implementation achieves almost full utilization of the available resources as long as there are unprocessed tokens to fill the cores. However, there are various long-running IBP coefficients, including one
  which decomposes in about $68$ hours and thus dominates the overall
  run-time. As a result, this form of parallelism is exhausted in the above mentioned computation and increasing the number of cores cannot lead to a further speedup.

Therefore, we use a second layer of parallelism available for the improved Leinartas algorithm (and described in Sections~\ref{sec:parallel} and \ref{Petrinet}) to speed up the decomposition of the individual
 IBP coefficients. 
 This Petri-net based approach  (detailed in Figures~\ref{fig:parallel_pfd} and \ref{fig:parallel_merge}) can decompose the most complicated coefficient in less than $12$ hours. We remark that the best achieved with standard task-based parallelism in \textsc{Singular} was about $19$ hours.  Applying the Petri net based parallelization with two layers of parallelism, {\sc pfd-parallel} finishes the decomposition for the complete set of IBP coefficients in about
$22$~hours and $15$
minutes on $520$ cores.\footnote{We expect that increasing the resources for the computation will lead to a further speedup, limited by the performance of the parallel computation of the most difficult coefficient. }

Table \ref{compression table} shows the size of the IBP coefficients
before and after applying the partial fraction decomposition, specified for different
numerator degrees of the Feynman integral. We observe that the compression ratio for the IBP
reduction coefficients increases with
the numerator degree.
\begin{table}[h]
\caption{Compression ratio of UT basis IBP reduction coefficients of
  the ``double-pentagon'' integral with different numerator degrees}\medskip
\label{compression table}
\centering
\ra{1.3}
\label{tab-marks}
  \begin{tabular}{@{}ccccc@{}}
  \toprule
degree & original size & compressed size & compression ratio\\
\midrule
2 & 73.1KB & 52.4KB & 1.4\\
3 & 23.6MB & 1.69MB & 14\\
4 & 1.19GB & 20.7MB & 59\\
5 & 18.4GB & 163MB & 115\\
total & 19.6GB & 186MB & 108\\
  \bottomrule
\end{tabular}
\end{table}

\begin{figure}[htbp]
    \caption{A scatterplot of the output file sizes against the input file sizes for the IBP reduction coefficients for the five-point non-planar ``double pentagon'' diagram for the degree up to $5$.}
    \label{fig:xbdeg_inout_sizes}
    \centerline{\includegraphics[height=8cm]{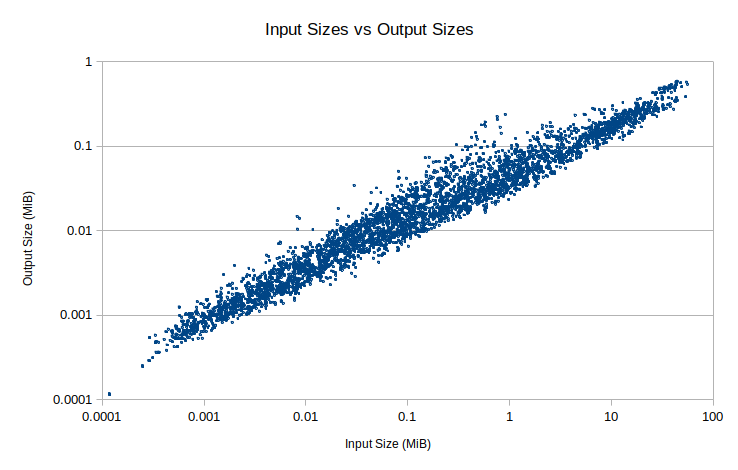}}
    \centering
  \end{figure}

\begin{figure}[htbp]
    \caption{A scatterplot of the compression ratios against the input file sizes for the IBP reduction coefficients for the five-point non-planar ``double pentagon'' diagram for the degree up to $5$.}
    \label{fig:xbdeg_compression}
    \centerline{\includegraphics[height=8cm]{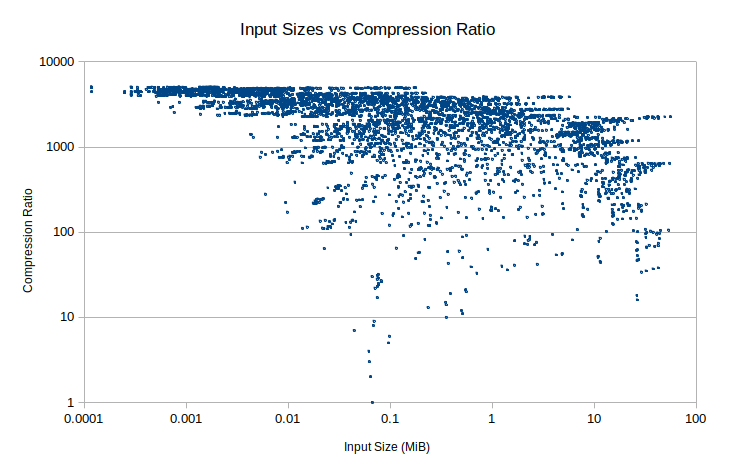}}
    \centering
  \end{figure}

Figure \ref{fig:xbdeg_inout_sizes} shows a scatterplot of the output file sizes against the input file sizes, and Figure \ref{fig:xbdeg_compression} details the compression ratios vs the input file sizes.
The final result for the IBP coefficients in the setting of this example can be
downloaded from
\begin{center}
\url{https://www.dropbox.com/s/v64rp4saphi1czh/double_pentagon_deg5_IBP.tar.gz}
\end{center}
containing a file with a $47\times 108$ matrix where each row corresponds to a target integral being reduced and each column corresponds to an UT integral in
the basis. The UT basis and the target integrals are also provided under this link.


\subsection{Simplifying analytic multi-loop scattering amplitudes}

Our program can in the same way be used to simplify analytic 
coefficients in the 
results of multi-loop scattering
amplitude computations. In this section, we consider the complicated coefficients arising in the result of \cite{Badger:2022ncb} on the two-loop leading colour helicity amplitudes for 
$W\gamma+j$ production at the LHC.

Our goal is to simplify the most complicated
coefficients  of the finite
remainder in ref.~\cite{Badger:2022ncb}. These coefficients explicitly are $452$
rational functions in terms of the kinematic variables,
\begin{equation}
  \label{eq:1}
  s_{12}, s_{123}, s_{23}, s_{234}, s_{34}, s_{56}\,.
\end{equation}
The denominators of the rational functions are factored. Among them, there are 20 linear denominator factors as
\begin{align}
&s_{12},s_{23},s_{12}+s_{23}-s_{123},s_{23}-s_{234},s_{12}-s_{34}-s_{123},\\
&s_{12}+s_{23}-s_{34}-s_{123},-s_{34}+s_{234},s_{12}-s_{34}+s_{234},s_{34},\\
&s_{23}+s_{34},s_{34}-s_{234},s_{23}+s_{34}-s_{234},-s_{56}+s_{123},-s_{56}+s_{234},\\
&s_{12}-s_{56}+s_{234},-s_{23}-s_{56}+s_{123}+s_{234},s_{12}-s_{34}-s_{56}+s_{234},\\
&s_{56}-s_{234},s_{23}+s_{56}-s_{123}-s_{234},-s_{12}+s_{34}+s_{56}-s_{234},
\end{align}
12 quadratic denominator factors as
\begin{align}
&s_{12} s_{56}-s_{12} s_{123}+s_{34} s_{123}-s_{56} s_{123}+s_{123}^2,\\
&s_{12} s_{123}-s_{34} s_{123}+s_{56} s_{123}-s_{123}^2-s_{12} s_{56},\\
&s_{123} s_{234}-s_{23} s_{56},s_{23} s_{56}-s_{123} s_{234},\\
&s_{12}^2-2 s_{12} s_{34}+s_{34}^2+s_{23} s_{56}-s_{12} s_{123}+s_{34} s_{123}+s_{12} s_{234}\nonumber\\&\quad-s_{34} s_{234}-s_{123} s_{234},\\
&s_{34} s_{234}+s_{123} s_{234}-s_{12} s_{234}-s_{23} s_{56}-s_{34} s_{56},\\
&s_{34} s_{234}+s_{56} s_{234}-s_{12} s_{234}-s_{234}^2-s_{34} s_{56},\\
&s_{12} s_{23}-s_{23} s_{34}-s_{23} s_{56}-s_{34} s_{56}-s_{12} s_{234}+s_{23} s_{234}+s_{34} s_{234}\nonumber\\&\quad+s_{56} s_{234}-s_{234}^2,\\
&s_{34} s_{56}+s_{12} s_{234}+s_{23} s_{234}-s_{34} s_{234}-s_{123} s_{234},\\
&s_{23} s_{56}+s_{34} s_{56}+s_{12} s_{234}-s_{34} s_{234}-s_{123} s_{234},\\
&s_{34} s_{56}+s_{12} s_{234}-s_{34} s_{234}-s_{56} s_{234}+s_{234}^2,\\
&s_{12}^2-2 s_{12} s_{34}+s_{34}^2-2 s_{12} s_{56}-2 s_{34} s_{56}+s_{56}^2,
\end{align}
and a degree-4 denominator factor as
\begin{align}
&s_{12}^2 s_{23}^2-2 s_{12} s_{23}^2 s_{34}+s_{23}^2 s_{34}^2-2 s_{12} s_{23}^2 s_{56}-4 s_{12} s_{23} s_{34} s_{56}-2 s_{23}^2 s_{34} s_{56}\nonumber\\&
+s_{23}^2 s_{56}^2+2 s_{12} s_{23} s_{34} s_{123}-2 s_{23} s_{34}^2 s_{123}+2 s_{23} s_{34} s_{56} s_{123}+s_{34}^2 s_{123}^2\nonumber\\&
-2 s_{12}^2 s_{23} s_{234}+2 s_{12} s_{23} s_{34} s_{234}+2 s_{12} s_{23} s_{56} s_{234}+2 s_{12} s_{23} s_{123} s_{234}\nonumber\\&
+2 s_{12} s_{34} s_{123} s_{234}+2 s_{23} s_{34} s_{123} s_{234}-2 s_{23} s_{56} s_{123} s_{234}-2 s_{34} s_{123}^2 s_{234}\nonumber\\&
+s_{12}^2 s_{234}^2-2 s_{12} s_{123} s_{234}^2+s_{123}^2 s_{234}^2
\end{align}


We observe that the representation of the functions can be significantly simplified by our {\sc pfd-parallel} package, thus reducing
the size of the data by a large factor. We rely on our implementation of the \textsc{MultivariateApart} algorithm, since 
the Leinartas algorithm does not perform well with regard to compression ratio and hence computation time in this example. We remark, that an automatism is under development to run different algorithmic strategies competitively using a wait-first parallelism to automatically obtain the best possible computation time or compression factor.



The computation was run on a single node with
48 CPU cores and 385~GB of RAM.
The byte sizes of the input and output, along with the total runtime and
compression ratio, are shown in Table
\ref{table:input_output_sizes_pfd2}.

\begin{table}[h]
  \centering
  \caption{Input and output sizes of the rational functions in  the two-loop leading
colour helicity amplitudes for $W\gamma+ j$ production. Here "ssi" means the data file format of {\sc Singular}. The words "indexed text" means results with abbreviated names of irreducible factors.}
  \label{table:input_output_sizes_pfd2}
  \ra{1.3}
  \begin{tabular}{@{}rrcrrrcrcr@{}}

    \toprule

    \multicolumn{2}{c}{Input size (MB)} &&
    \multicolumn{3}{c}{Output size (MB)} &&

    Time &&
    \multirow{2}{*}{\shortstack[r]{Compression\\ ratio}} \\

    \cmidrule{1-2} \cmidrule{4-6}
    text & ssi & & ssi & text & indexed text  && && \\

    \cmidrule{1-2} \cmidrule{4-6} \cmidrule{8-8} \cmidrule{10-10}

    1096.76 & 392.30 && 9.20 & 9.24 & 5.85 && 5h34m && 187.60 \\

    \bottomrule

  \end{tabular}
\end{table}

\begin{figure}[htbp]
\centerline{\includegraphics[height=8cm]{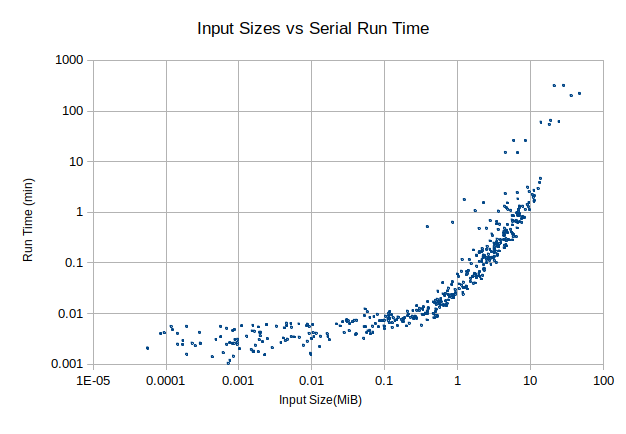}}
\caption{Scatterplot of computation times vs input byte size for the two-loop leading colour helicity amplitudes for 
$W\gamma+j$ production.}
\label{graph:pfd_scatter_sizes}
\end{figure}

\begin{figure}[htbp]
\centerline{\includegraphics[height=8cm]{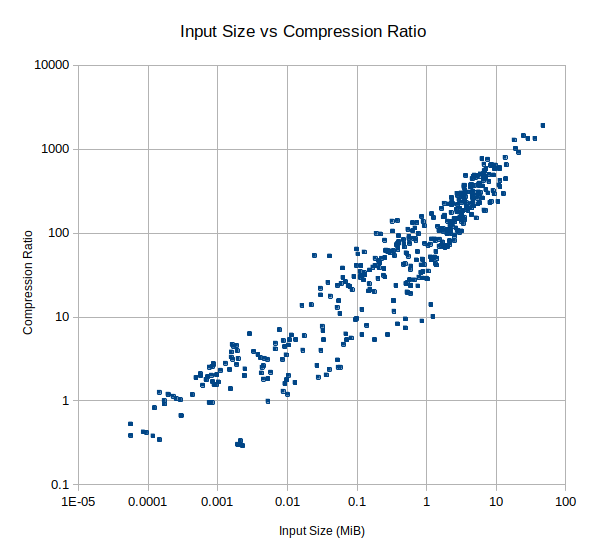}}
\caption{Scatterplot the compression ratio vs the input byte size for the two-loop leading colour helicity amplitudes for 
$W\gamma+j$ production.}
\label{graph:pfd_scatter_compression}
\end{figure}

The indexed text output byte size vs the text input byte size is illustrated as a scatter plot in Figure~\ref{graph:pfd_scatter_sizes}.
The the  compression ratio vs the text input byte size (for the
indexed output format) are shown in a similar fashion in Figure
\ref{graph:pfd_scatter_compression}. The output result after the
partial fraction, can be downloaded from,
\begin{quotation}
 \url{https://www.dropbox.com/s/xgfj6x1qg25n44w/pfd_W_FiniteRemainder.tar.gz}
\end{quotation}
.

\begin{figure}[htbp]
\centerline{\includegraphics[height=8cm]{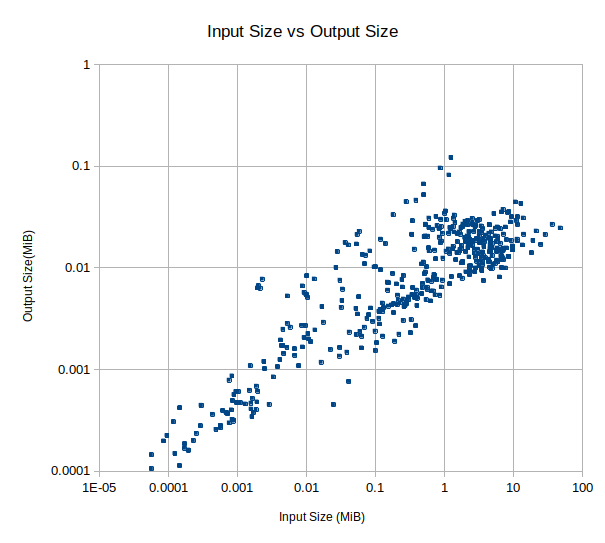}}
\caption{Scatterplot of output byte size vs input byte size for two-loop leading colour helicity amplitudes for 
$W\gamma+j$ production.}
\label{graph:pfd_scatter_sizes}
\end{figure}

\section{Summary}
\label{sec:outlook}
In this paper, we present a massively parallel implementation of the
multivariate partial fraction algorithms from~\cite{Boehm:2020ijp}
and~\cite{Heller:2021qkz}. 
Our implementation relies only on open source software. Based on  the \textsc{Singular}/\textsc{GPI-Space} 
framework for parallel computations in computer algebra, it combines parallelisation over different target
coefficients, and a parallel decomposition of individual coefficients. Using this approach, problems where individual coefficients dominate the overall run-time can be handled in an efficient way. In the paper, we demonstrate the power of our package
by cutting-edge examples: (1) simplifying very complicated IBP
reduction coefficients (2) rational function coefficients in
analytic multiloop scattering amplitudes. We expect that our package can be of essential
use in multiloop amplitudes computations.

The workflow of multivariate partial fraction computation with several levels of parallelism
is realized in an efficient way using the  workflow management system \textsc{GPI-Space}. The approach is based on the idea of separation of coordination and computation, and models the fundamental algorithmic structure in form of a
Petri net.  This allows for automated parallelisation, makes the code easily maintainable and suitable for transparently incorporating future improvements. The Petri net based 
parallelisation methods developed in this paper are expected to be useful
in a much broader sense, for example, for  applications in modular 
computations. Respective code is under development.


\section*{Acknowledgement}
We acknowledge Taushif Ahmed, Christoph Dlapa, Bo Feng, Alessandro Georgoudis, Matthias Heller,
Johannes Henn, Zhao Li, Yanqing Ma, Hans Sch\"onemann and Huaxing
Zhu for very useful
discussions. In particular, we thank  Matthias Heller and Andreas von Manteuffel for the \textsc{MultivariateApart}
algorithm ~\cite{Heller:2021qkz}. They kindly
gave us the permission to implement this algorithm as part of our package powered by the
\textsc{Singular}/\textsc{GPI-Space} framework. We also deeply thank Simon
Zoia, for providing us the rational function coefficients in the work \cite{Badger:2022ncb}
(the two-loop leading colour helicity amplitudes for 
$W\gamma+j$ production at the LHC) to test the efficiency of our
multivariate partial fractioning framework.

The work of YZ was supported from the NSF of China through Grant
No. 11947301, 12047502 and No. 12075234. 
Gef\"ordert durch die Deutsche Forschungsgemeinschaft (DFG) - Projektnummer 286237555 - TRR 195 \linebreak (Funded by the Deutsche Forschungsgemeinschaft (DFG, German Research Foundation) - Project-ID 286237555 - TRR 195). The work of JB and MW was supported by Project B5 of SFB-TRR 195. The work of LR was supported by Project A13 of SFB-TRR 195 and Potentialbereich \emph{SymbTools - Symbolic Tools in Mathematics and their Application} of the Forschungsinitative Rheinland-Pfalz.


\bibliographystyle{elsarticle-num}
\bibliography{bibtex}

\end{document}